\documentclass[a4paper,fleqn,usenatbib]{mnras}   %mn2e} %{mnras}

\usepackage[T1]{fontenc}
\usepackage{ae,aecompl}
\usepackage{graphicx}	% Including figure files
\usepackage{amsmath}	% Advanced maths commands
\usepackage{amssymb}	% Extra maths symbols

\newcommand{\civalone}{\textrm{C}\textsc{iv}}
\newcommand{\ciiialone}{\textrm{C}\textsc{iiii}]}

\newcommand{\hi}{\textrm{H}\textsc{i}}

\newcommand{\ha}{\ifmmode {\rm H}\alpha \else H$\alpha$\fi}
\newcommand{\hb}{\ifmmode {\rm H}\beta \else H$\beta$\fi}
\newcommand{\lya}{\ifmmode {\rm Ly}\alpha \else Ly$\alpha$\fi}
\newcommand{\pg}{\ifmmode {\rm P}\gamma \else Pa$\gamma$\fi}
\newcommand{\lyb}{\ifmmode {\rm Ly}\beta \else Ly$\beta$\fi}
\newcommand{\lyg}{\ifmmode {\rm Ly}\gamma \else Ly$\gamma$\fi}
\newcommand{\ciii}{\textrm{C}\textsc{iii}]\ensuremath{\lambda1908}}
\newcommand{\ciiidoub}{\textrm{C}\textsc{iii}]\ensuremath{\lambda\lambda1907,1909}}

\newcommand{\nv}{\textrm{N}\textsc{v}\ensuremath{\lambda1240}}

\newcommand{\civ}{\textrm{C}\textsc{iv}\ensuremath{\lambda1548,1550}}

\newcommand{\heii}{\textrm{He}\textsc{ii}\ensuremath{\lambda1640}}
\newcommand{\oiiiuv}{\textrm{O}\textsc{iii}]\ensuremath{\lambda1661,1666}}

\newcommand{\flyc}{\ifmmode  \mathrm{f}_\mathrm{esc}\mathrm{(LyC)} \else $\mathrm{f}_\mathrm{esc}\mathrm{(LyC)}$\fi}

\def\kms{km s$^{-1}$}

\def\ergs{\ifmmode \mathrm{erg\hspace{1mm}s}^{-1} \else erg s$^{-1}$\fi}
\def\ergscm{erg s$^{-1}$ cm$^{-2}$}
\def\micron{\ifmmode \mu\mathrm{m} \else $\mu$m\fi}
\def\msun{\ifmmode \mathrm{M}_{\odot} \else M$_{\odot}$\fi}
\def\msunyr{\ifmmode \mathrm{M}_{\odot} \hspace{1mm}{\rm yr}^{-1} \else $\mathrm{M}_{\odot}$ yr$^{-1}$\fi}
\def\zsun{\ifmmode Z_{\odot} \else Z$_{\odot}$\fi}
\def\lsun{\ifmmode L_{\odot} \else L$_{\odot}$\fi}
\def\mstar{\ifmmode \mathrm{M}_{\star} \else M$_{\star}$\fi}

\title[Star Cluster Formation Caught in the Act at z=6] %[Dense high redshift dwarf galaxies]
{Massive Star cluster formation under the microscope at z=6}

\author [E.~Vanzella et al.]{
\parbox[t]{\textwidth}{E.~Vanzella$^1$\thanks{E-mail: eros.vanzella@oabo.inaf.it},
F.~Calura$^1$, M.~Meneghetti$^1$, M.~Castellano$^2$, G.~B.~Caminha$^3$, A.~Mercurio$^4$, G. Cupani$^{5}$, P.~Rosati$^{1,6}$, C.~Grillo$^{7}$, R.~Gilli$^1$, M.~Mignoli$^1$, G.~Fiorentino$^1$, C.~Arcidiacono$^1$, M.~Lombini$^1$ and F.~Cortecchia$^1$
}
\vspace*{8pt}\\
$^1$INAF -- OAS, Osservatorio di Astrofisica e Scienza dello Spazio di Bologna, via Gobetti 93/3, I-40129 Bologna, Italy\\ %INAF -- Osservatorio Astronomico di Bologna, via Gobetti 93/3, 40129 Bologna, Italy\\
$^2$INAF -- Osservatorio Astronomico di Roma, Via Frascati 33, I-00078 Monte Porzio Catone (RM), Italy\\
$^3$Kapteyn Astronomical Institute, University of Groningen, Postbus 800, 9700 AV Groningen, The Netherlands\\
$^4$INAF -- Osservatorio Astronomico di Capodimonte, Via Moiariello 16, I-80131 Napoli, Italy\\
$^5$INAF -- Osservatorio Astronomico di Trieste, via G. B. Tiepolo 11, I-34143, Trieste, Italy\\
$^6$Dipartimento di Fisica e Scienze della Terra, Universit\`a degli Studi di Ferrara, via Saragat 1, I-44122 Ferrara, Italy\\
$^7$Dipartimento di Fisica, Universit\`a  degli Studi di Milano, via Celoria 16, I-20133 Milano, Italy\\
}

\begin{document}
\date{}
\maketitle

\begin{abstract}
We report on a superdense star-forming region with an effective radius ($R_e$) smaller than 13 pc identified at z=6.143
and showing a star-formation rate density $\Sigma_{SFR} \sim 1000$ M$_{\odot}$~yr$^{-1}$~kpc$^{-2}$
(or conservatively $>300$ M$_{\odot}$~yr$^{-1}$~kpc$^{-2}$).
Such a dense region is detected with S/N $\gtrsim40$ hosted by a dwarf extending over 440 pc, dubbed D1.
D1 is magnified by a factor 17.4($\pm5.0$) behind the Hubble Frontier Field galaxy cluster
MACS~J0416 and elongated tangentially by a factor $13.2\pm4.0$ (including the systematic errors).  
The lens model accurately reproduces the positions of the confirmed multiple images
with a r.m.s. of $0.35''$. 
D1 is part of an interacting star-forming complex extending over 800 pc.
The SED$-$fitting, the very blue ultraviolet slope ($\beta \simeq -2.5$, F$_{\lambda} \sim \lambda^{\beta}$)
and the prominent \lya\ emission of the stellar complex imply that very young ($< 10-100$Myr), moderately
dust-attenuated (E(B-V)<0.15) stellar populations are
present and organised in dense subcomponents.
We argue that D1 (with a stellar mass of $2\times10^{7}$ \msun) might contain a
young massive star cluster of M $\lesssim 10^{6}$ \msun\ and M$_{UV} \simeq -15.6$ (or $m_{UV}=31.1$),
confined within a region of 13 pc, and not dissimilar from some local super star clusters (SSCs). The ultraviolet
appearance of D1 is also consistent with a simulated local dwarf hosting a SSC placed at z=6 and lensed back to the observer.
This compact system  fits into some popular globular cluster formation scenarios.
We show that future high spatial resolution imaging (e.g., E$-$ELT/MAORY-MICADO and VLT/MAVIS) 
will allow us to spatially resolve light profiles of 2-8 pc.
\end{abstract}


\section{Introduction}
The observational investigation of star-formation at high redshift ($z\gtrsim6$)
at very small physical scales (at the level of star-forming complexes of $\lesssim 200$ pc
including super star clusters) is a new challenge in observational cosmology
\citep[e.g.,][]{rigby17,joh17,livermore15,vanz17b,vanz17c,mirka17,MA18,cava18}.
Thanks to strong gravitational lensing,
 the possibility to catch and study globular clusters precursors (GCP) is becoming a real fact,
both with statistical studies \citep[e.g.,][]{renzini17,boylan18,vanz17b,elme12} and by inferring the physical properties of individual objects  \citep[e.g.,][]{vanz17b,vanz17c}.
The luminosity function of forming GCs has also been addressed for the first
time \citep{bouwens18,boylan18} and their possible contribution to the ionising background 
is now under debate \citep[e.g.,][]{ricotti02, schaerer11,katz13,boylan18}.
While still at the beginning, the open issues of GC 
formation \citep[e.g.,][]{bastian17, renzini15,renaud18} and what sources caused 
reionization \citep[e.g.,][]{robertson15,yue14} can be addressed with the same observational approach, 
at least from the high-z prospective. 
This is a natural consequence of the fact that the search for extremely
faint sources possibly dominating the ionising background \citep[e.g.,][]{robertson15,finkelstein15,bouwens16a,bouwens16b, alavi16,yue14,dayal18} 
plausibly matches the properties a GCP would have both in terms of
stellar mass and luminosity \citep[e.g.,][]{renzini17,schaerer11,boylan18,bouwens18}
and this eventually depends on the different GCP formation scenarios
\citep{bastian17,renzini17,renzini15,zick18,kim18,ricotti16,li17}.  
A way to access low-luminosity regimes $-$ otherwise not attainable in the blank fields $-$ is by exploiting
gravitational lenses. Other than ``simply'' counting objects at unprecedented flux limits,
the strong lensing amplification allow us to probe the structural parameters down to the scale of a few tens of parsec
\citep[e.g.,][]{rigby17,livermore15,vanz17b,vanz17c,kawamata15} and witness clustered star-forming regions and/or
star clusters otherwise not spatially resolved in non-lensed field studies. 
The lens models are subjected to a strict validation thanks to dedicated 
simulations and observational campaigns with Hubble  \citep[e.g.,][]{meneghetti17, atek18} in conjunction to
unprecedented (blind) spectroscopic confirmation of hundreds of multiple images with 
VLT/MUSE\footnote{{\it www.eso.org/sci/facilities/develop/instruments/muse.html} \citep[][]{bacon10,bacon15}}
in the redshift range $3<z<6.7$  \citep[e.g.,][]{karman17,cam17a,cam17b,mahler18}. Such analyses are providing
valuable insights on the systematic errors on magnification maps. In some (not rare) conditions
the uncertainty on large magnification $\mu > 10$ can be significantly lowered to a few percent 
by exploiting the measured relative fluxes among multiple images that 
provide an observational constraint on the relative magnifications
\citep[e.g.,][]{vanz17b,vanz17c}.
These methods allow us to determine the absolute physical quantities, like the luminosity, sizes, stellar mass,
and star-formation rates with uncertainties not dominated by the aforementioned systematics.

\begin{figure*}
\centering
\includegraphics[width=17cm]{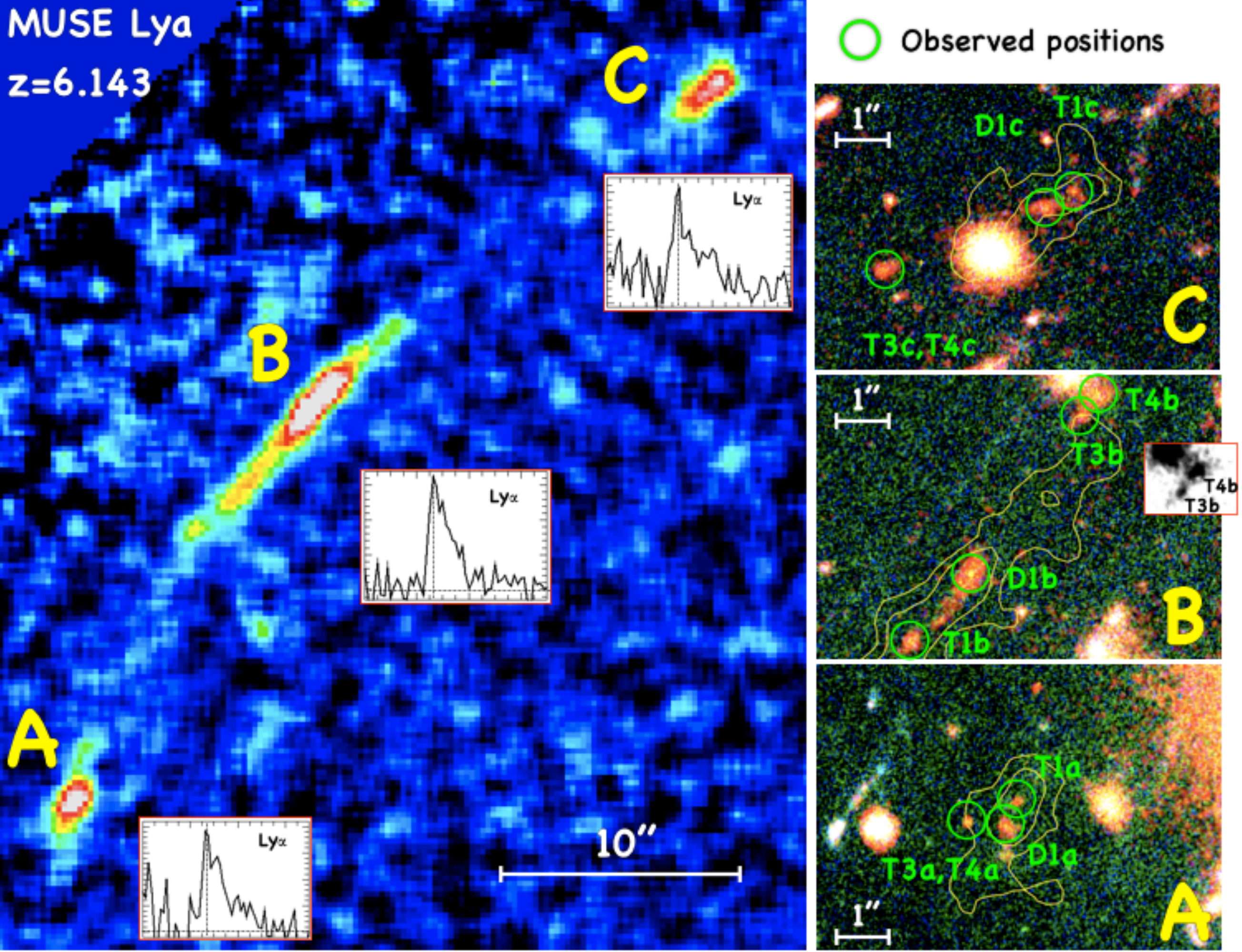}  %Fig_MULTIPLE3.pdf}
\caption{{\it Left}: The wide  \lya\ arc ($50''$) at z=6.143 observed with MUSE and 
weighted-averaged over 12 slices ($\Delta v \simeq 500$\kms). Three multiple images are indicated (A, B and C, see \citealt{cam17a}) with their
associated \lya\ lines extracted from the MUSE spectra. The image B is the most magnified among the
three and studied in detail in the present work. On the right side, the three panels from top to bottom show the
zoomed regions in the colour HST image (red channel = F105W, green channel = F814W and blue channel= F606W) of the main images A, B and C, including
the observed positions (indicated with green circles) of the multiple images of relevant objects (D1, T1 and the T3-T4 pair).
The inset in the middle-right panel is the F105W showing the double knot morphology of T3$-$T4, which is barely detected in the less 
magnified counter images A and C.  
The yellow contours show the MUSE \lya\ emission at 3 and 7 sigma level.}
\label{multiple}
\end{figure*}

A more complex issue is related to 
the role of such a nucleated star formation 
on the ionisation of the surrounding medium, eventually leaking into the intergalactic medium.
Probing the presence of optically thin (to Lyman continuum) channels
or cavities which cause the ionising leakage from these tiny sources \citep[e.g.,][]{calu15,behrens14} 
will represent the next challenge.
The presence of diffuse \lya\ emission (observed as nebulae or halos or simply offset emissions)
often detected around faint sources may provide a first route to address this issue (e.g., \citealt{cam16,vanz17a,lel17}, see 
also \citealt{gallego18}), along with the recent detection of ultraviolet high$-$ionisation nebular lines
like \civ, \heii, \oiiiuv\ or \ciiidoub\ suggesting that hot stars and/or nuclear 
contribution might be present, making some sources highly efficient Lyman continuum emitters \citep[e.g.,][]{stark14,stark15a,stark15b,stark17,vanz17c}.
However, the final answer, especially at $z>3-6$, will be addressed only with JWST by monitoring the spatial
distribution of the Balmer lines, and possibly look for induced fluorescence by  the Lyman continuum leakage
up to the circum galactic medium and/or to larger distances, i.e., the IGM \citep[e.g.,][]{ribas17}.

While giant ultraviolet clumps have been studied at high redshift \citep[e.g.,][]{cava18, elme13,guo12,forst11,genzel11}, the 
direct observation of young star clusters at cosmological distances is challenging. 
Given the typical HST pixel scale ($0.03''/$pix) and spatial sampling (e.g., $0.18''$ FWHM of the PSF in the WFC3/F105W
band), the most stringent upper limit on the physical size attainable after a proper 
PSF-deconvolution\footnote{As can be performed with {\tt Galfit}, see simulations reported in \cite{vanz16a,vanz17b} and discussion in 
\citealt{peng10}.} is 168(84) pc, corresponding to 1.0(0.5) pixels at redshift 6, in a non-lensed field.
If compared to the typical effective radii of local young massive clusters of 
$R_e < 20$ pc~\footnote{$R_e \simeq 1-8$ pc for masses of the clusters of $< 10^6$\msun,  \citealt[e.g.,][]{ryon17}, 
or slightly larger radii, $<20$ pc,  for those more massive, $> 10^6$\msun\ and identified in merging galaxies, 
\citealt[][]{port10,linden17}.},  assuming this value holds also at z=6,
it becomes clear why
strong lensing is crucial if one wants to approach such a scale with HST.
As shown in \citet{vanz17b,vanz17c} the lensing magnification can significantly stretch the image
along some preferred direction (up to a factor 20, tangentially or radially with respect to the lens) 
allowing us to probe the aforementioned small sizes of $10-30$ pc. 
This effect was exploited in a study of a sample of objects behind the Hubble Frontier Field galaxy cluster
MACS~J0416 (\citealt{vanz17b}, see also \citealt{bouwens18}). 

The identification of a very nucleated (or not spatially resolved) object
despite a large gravitational lensing stretch is an ideal case
where to search for single stellar clusters (and potential GCP). 
Here we report on such a case and perform new analysis on a pair of objects already presented
in \citet{vanz17b} but with significantly improved size measurements, refined lensing modelling and 
SED$-$fitting. The objects discussed in this work, dubbed D1 and T1 at z=6.143, 
correspond to D1 and GC1 previously reported by \citet{vanz17b}.
The combination of the main physical quantities like the star formation rate and the sizes reveals
an extremely large star-formation rate surface densities, 
lying in a poorly explored region of the Kennicut-Schmidt law \citep{kennicutt98b,bigiel10}.

In Sect.~2.1, 2.2 and 2.3 the refined lens model, ultraviolet morphology and the physical properties of 
the system are presented. Using the \lya\ properties and the SED-fitting results 
the emerging dense star-formation activity is discussed in Sect. 2.4. In Sect.~3 we simulate a local 
star-forming dwarf hosting a super-star cluster (NGC~1705) to z=6.1 and applying strong lensing.
In Sect.~4 we discuss the results and the identification of a super-star cluster at z=6.1, compared to
local young massive clusters. Sect.~5 summarises the main results. 
We assume a flat cosmology with $\Omega_{M}$= 0.3, $\Omega_{\Lambda}$= 0.7 and $H_{0} = 70$ km s$^{-1}$ Mpc$^{-1}$,
corresponding to 5660 physical parsec for 1 arcsec separation at redshift z=6.143. If not specified, the distances reported in the
text are physical.

\begin{figure*}
\centering
\includegraphics[width=17cm]{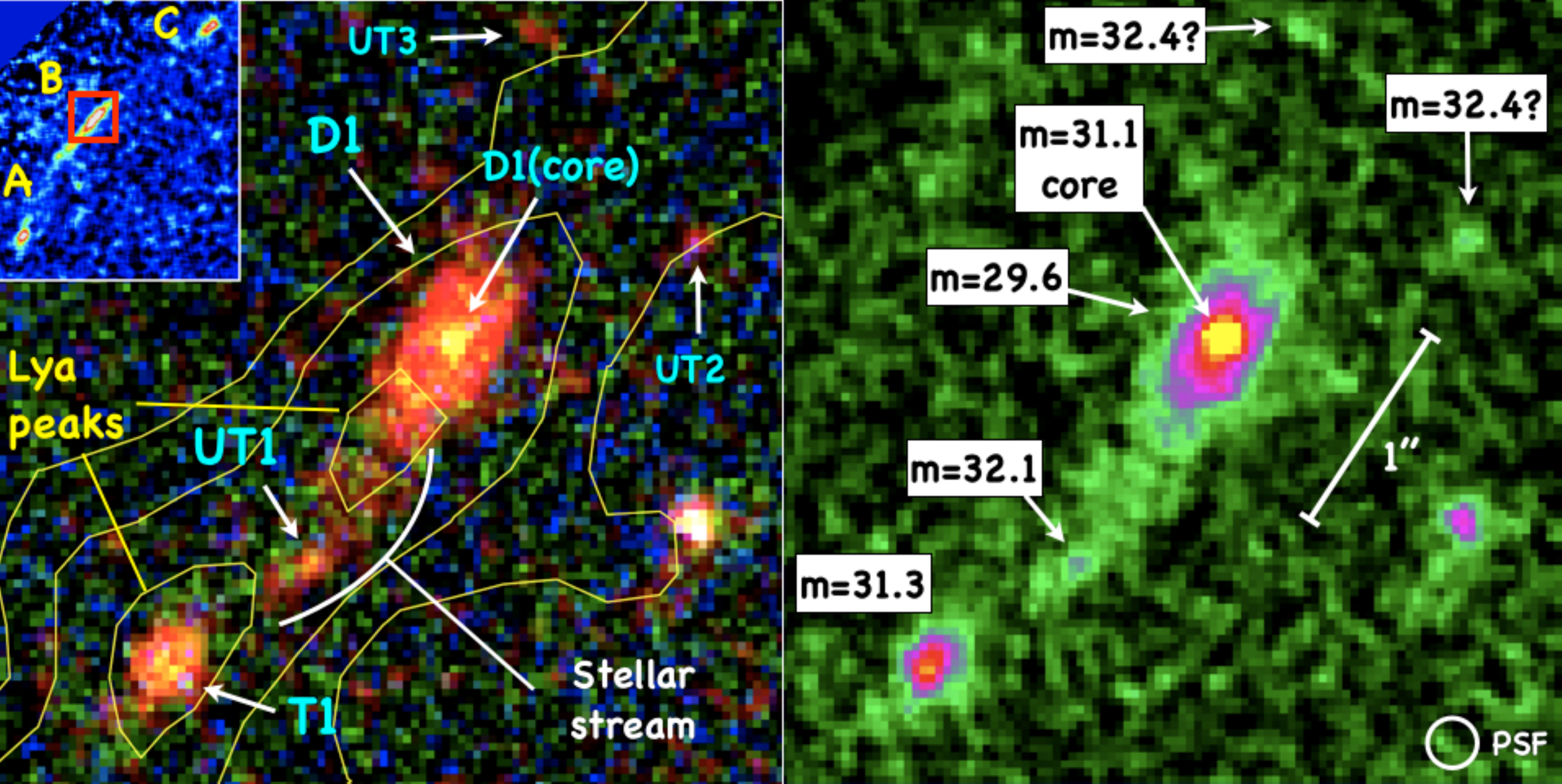} %Fig_ZOOM_D1T1.pdf}
\caption{Color composite (left) and the WFC3/F105W image (right) of the field under study containing the sources D1 and T1.
This region corresponds to the red square in the top-left inset
which shows the extended \lya\ arc from MUSE (see Figure~\ref{multiple}). 
Sources are labelled (left), along with their de-lensed 
F105W magnitudes (right). Note the prominent symmetric core of D1 despite the large
tangential magnification and the
presence of a stellar stream possibly connecting D1 and T1, also including a star-forming knot, dubbed UT1. Other faint knots are shown,
UT2 and UT3, with de-lensed magnitudes fainter than 32.
The HST F105W PSF is shown in the bottom right.}
\label{zoom}
\end{figure*}

\begin{figure*}
\centering
\includegraphics[width=17cm]{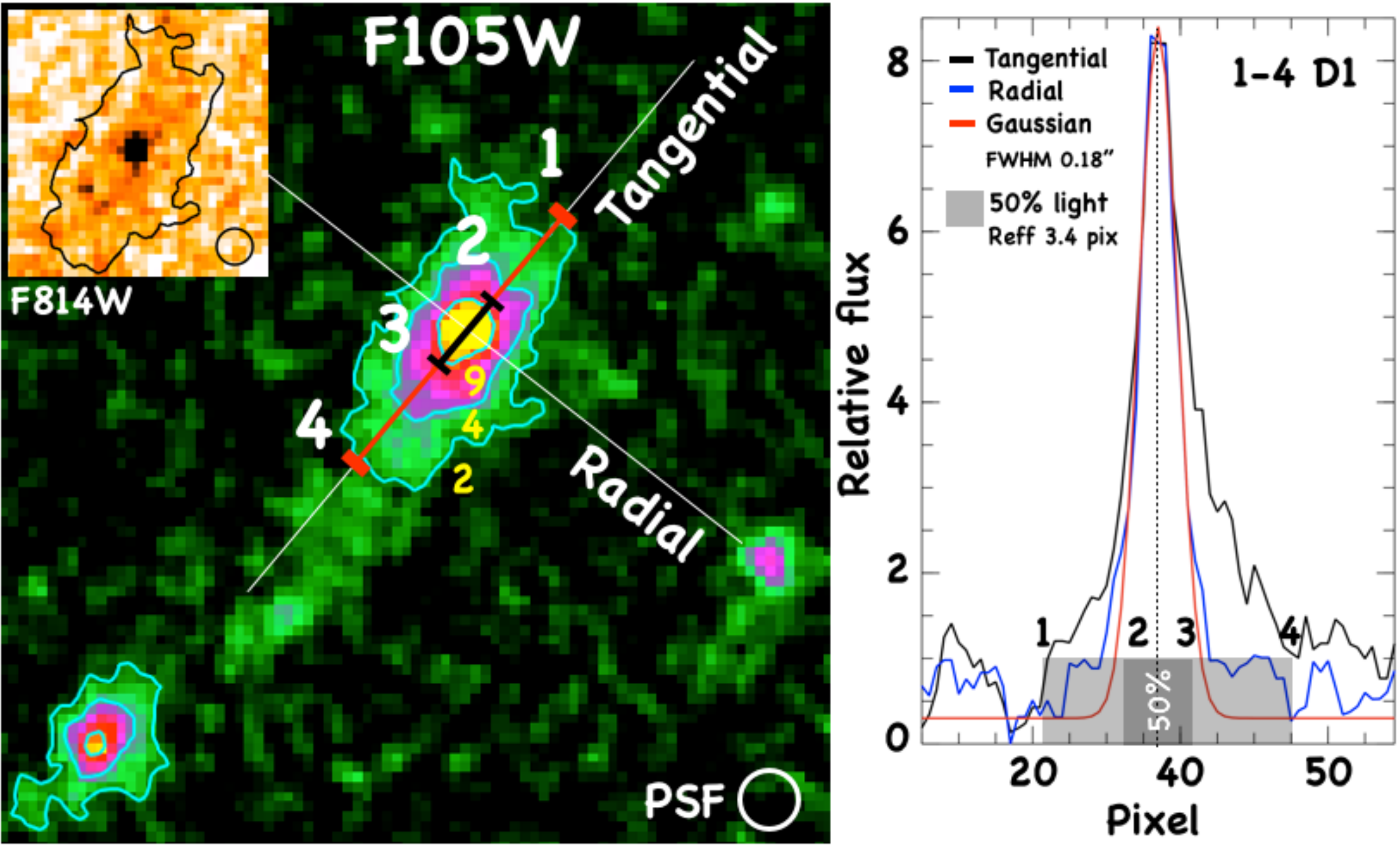} %Fig_Reff_EMPIRICAL.pdf}
\caption{The tangential and radial profiles (labelled) extracted from the F105W band and centred on the core of D1 are shown in the right panel. 
The Gaussian shape with FWHM of $0.18''$ (equal to the width of the PSF, or 6 pix) is also 
superimposed with a red line and is consistent with the plotted radial profile (as expected given the modest radial magnification, $\mu_R \simeq 1.3$).
The 50\% of the light along the tangential direction (marked with the segment 1-4 in the left panel) is enclosed within $\sim9$ pix 
(segment 2-3) and shown with a grey stripe, corresponding to a PSF-corrected size of $\sim 6.8$ pix (and a radius of 3.4 pix or $R_{e} \simeq 44$ pc).
In the left panel the F105W image of D1 and T1 is shown, in which the cyan contours mark the 2$\sigma$, 4$\sigma$ and 9$\sigma$ levels
above the background. The PSF size is also shown with a white circle ($0.18''$ diameter). In the top-left inset the 
F814W image of  D1 is also shown with the same 9$\sigma$ contour based on the F105W band. The presence of the 
IGM in the F814W band attenuates the signal, that, however, still reveals a nucleated emission (compatible with the F814W PSF, FWHM=$0.16''$).}
\label{re}
\end{figure*}

\section{Reanalysing the z=6.143 system in MACS~J0416}

\subsection{A robust lensing model}
\label{lensmodel}

In \citet{vanz17b} we used the lens model developed by \citet{cam17a} \citep[see also,][]{grillo15} to infer the intrinsic physical
and morphological properties of the system shown in Figure~\ref{multiple}, made by a star-forming complex
including the objects D1 and T1 (meaning Dwarf 1 and Tiny 1, respectively). We will refer in the following to
the system ``D1T1'' to indicate the entire system including the stellar stream connecting the two 
(see Figure~\ref{zoom}, in which much fainter sources, dubbed ``Ultra Tiny'' (UT),
UT1,2,3 are also indicated and mentioned in the discussion).
The model was tuned to reproduce the positions of more than 100 confirmed multiple images, belonging to 37 individual systems,
spanning the redshift range $3-6.2$.
Here we focus on the system at redshift 6.143 that recently has been further enriched by (at least) 13 individual objects
producing more than 30 multiple images all at  $z\simeq6$, some of them already spectroscopically confirmed at the same redshift of D1T1
(such an overdensity will be presented elsewhere) and others still based on robust photometric
redshifts \citep[e.g.,][]{castellano16b}. The new systems and the multiple
images are also consistent with the  expected positions predicted from the aforementioned lens model. 
An example is the system dubbed T3 and T4, a pair of sources showing the same colours and dropout signature
as D1T1 (this object is also present in the \citealt{castellano16b} catalog with zphot $\simeq 6$). Indeed, the lens model allows us to
reliably identify the multiple images, corroborating also the photometric redshift with the lens model itself. 
Figure~\ref{multiple} shows the well spatially resolved T3,T4 pair in the most magnified (tangentially$-$stretched) image B, while
the counter-images A and C appear as a single merged object (though image C still shows an elongation, as expected). 
These new identifications allow us to further tune and better constrain the lens model lowering the uncertainties 
of the magnification maps at z=6 (an ongoing deep MUSE AO-assisted program, 22h PI Vanzella,
will explore the redshifts of the overdensity). 
Figure~\ref{multiple} shows the 9 identified multiple images of the system D1T1 and T3$-$T4 (marked with green circles).
 The positions are reproduced with an r.m.s. of
$0.35''$. The same accuracy ($0.38''$) is measured even including the aforementioned $z=6$ structure using the
spectroscopically confirmed and/or robust photometric redshifts objects not utilised by \citet{cam17a} at the time the lens
model was constructed. This highlights the excellent predicting power and the reliability of the model on 27 multiple
images in total (9 individual objects at $z\simeq6$ not shown here, Vanzella et al. in preparation).

\begin{figure*}
\centering
\includegraphics[width=16.5cm]{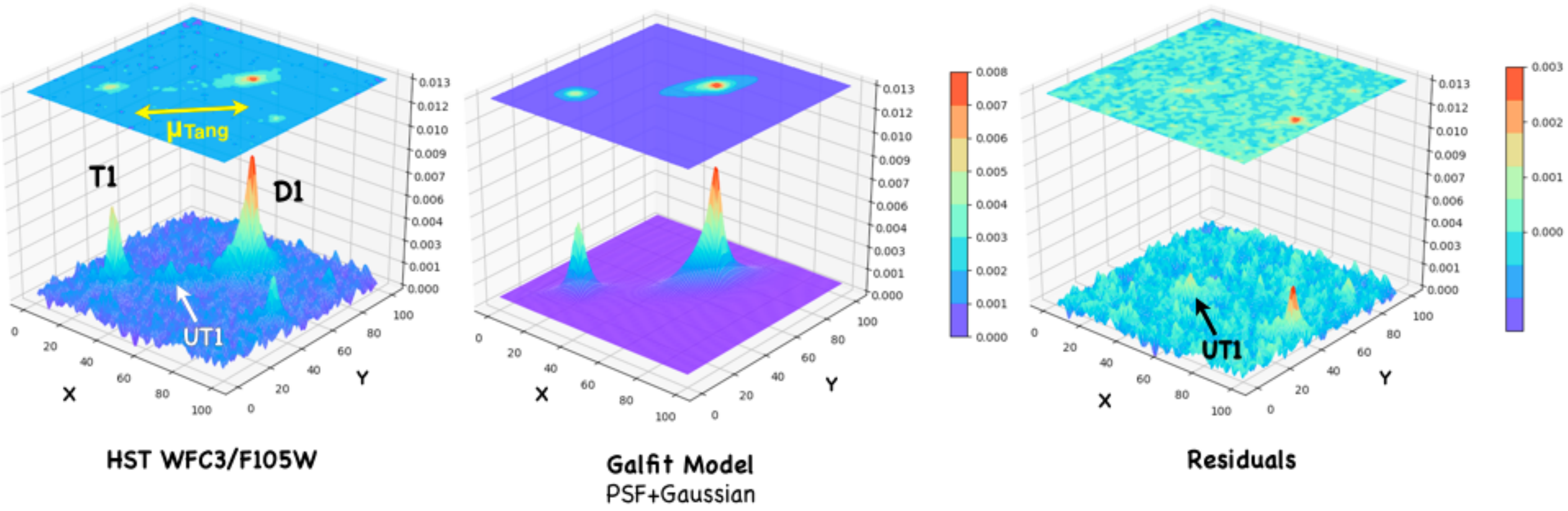} %Fig_Galfit_3D_2.pdf} %Fig_GALFIT.pdf}
\caption{{\tt Galfit} modelling in the F105W band. The best model includes two components: a diffuse
Gaussian component well reproducing the extended envelope with superimposed a pure PSF that well reproduces
the morphology of the core. From left to right: the HST original images, the {\tt Galfit} model and the residuals 
(i.e., the difference of the two quantities) are shown in the pixel domain in the XY-plane and colour-coded in the Z-axis 
following the HFF/WFC3 counts in the F105W band.}
\label{galfit}
\end{figure*}

As already discussed in \citet{vanz17b}, we probe extremely
small physical sizes in the z=6.143 system, exploiting the maximum magnification component, 
which is along the tangential direction in this case, as apparent from arc-like shape of the \lya\ emission (see Figure~\ref{multiple}). 
Table~\ref{tab} reports the total, $\mu_{tot}$, and tangential, $\mu_{tang}$,
magnifications at the positions of D1.  They are fully consistent with the previous estimates, but
the uncertainties are now decreased thanks to the additional constraints discussed above.
Statistical errors are of the order of 5\%. To access systematic errors we rely on the extensive 
simulations reported by \citet{meneghetti17}, aimed at performing an unbiased comparison among different
lens modelling techniques specifically applied to the Hubble Frontier Field project\footnote{{\it https://archive.stsci.edu/prepds/frontier/lensmodels/}} 
(including the code {\tt LENSTOOL} used by \citealt{cam17a}). 
In particular, the accuracy in reproducing the
positions of multiple images (e.g., r.m.s.) correlates with the total error on magnification,
especially at the position of the multiple images themselves  (Figure~26 of \citealt{meneghetti17}).
In the present case, considering the lens model adopted and the accuracy in reproducing  the positions
of the multiple images, 
the expected systematic uncertainty on the magnification factors is not larger than 30\%.
This translates to a 1-sigma error for the magnification on D1 of $\mu_{tot}=17.4 \pm 1_{stat} \pm 5_{syst}$,
and for the tangential magnification,  $\mu_{tang}=13.2 \pm 0.5_{stat} \pm 4_{syst}$.
The same arguments apply to the other compact source T1, for which we have 
$\mu_{tot}=24 \pm 2_{stat} \pm 7_{syst}$,
and  $\mu_{tang}=18 \pm 1_{stat} \pm 5_{syst}$ (see Table~\ref{tab}).

\begin{figure}
\centering
\includegraphics[width=8.5cm]{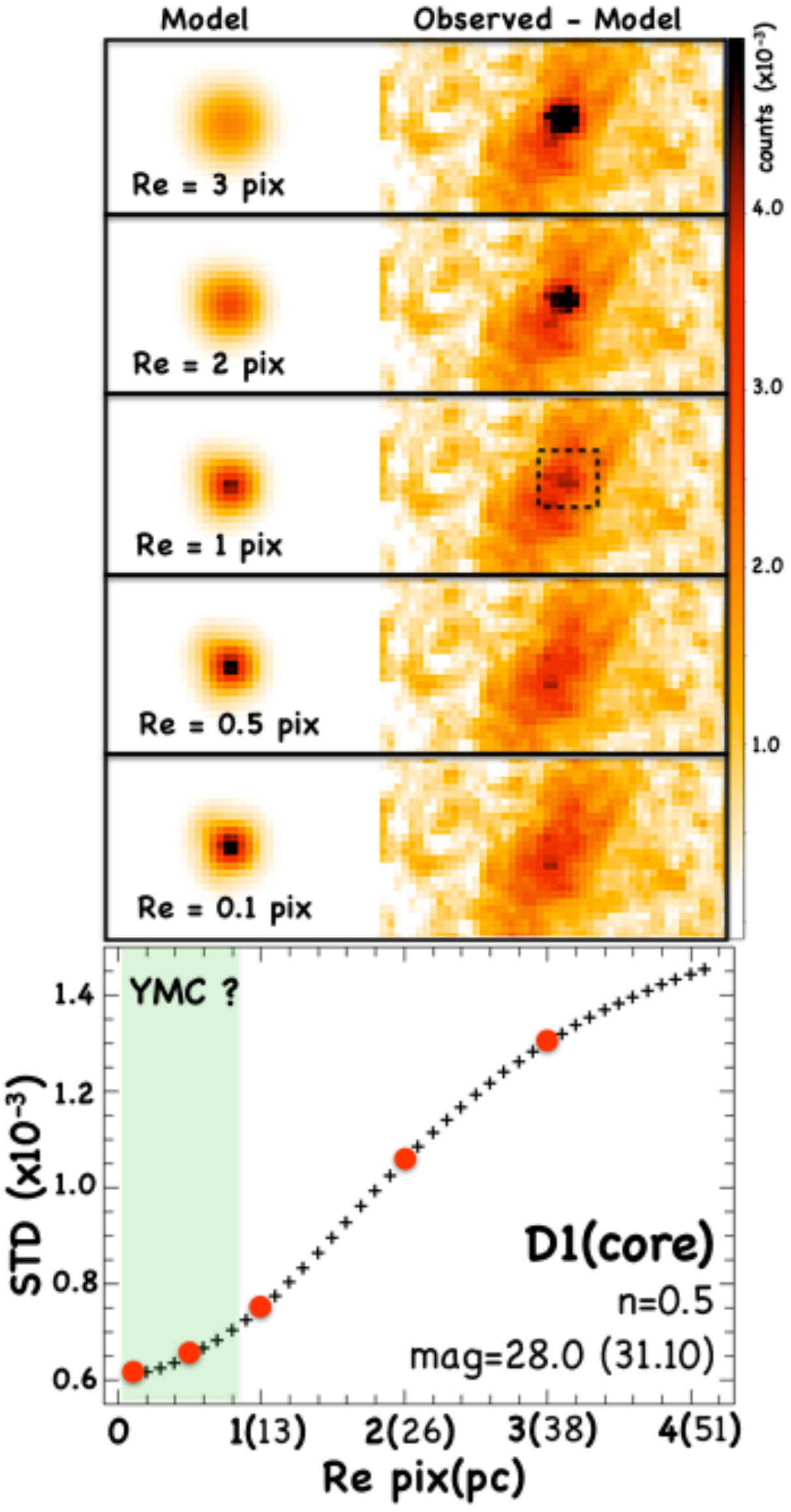} 
\caption{A detail of the {\tt Galfit} modelling of the D1 nuclear emission in the F105W band. 
The five images on the top show the models and residuals (observed - model) for various values of the effective radius ($R_e$). 
At $R_e = 1$ pix the residual is still positive, suggesting the object has a $R_e <1$ pix (see text for details). In the bottom panel, 
the behaviour of the standard deviation (calculated at each step within the area outlined with the dotted box shown in the panel at $R_e=1$ pix)
is shown with a finer grid of $R_e$ (at fixed $n=0.5$ and magnitude 28), and shows a monotonically decreasing shape
never reaching a minimum, 
suggesting that (at given the S/N ratio) the intrinsic size is not recovered and lies below the deconvolution capabilities of
the algorithm, implying an $R_e < 1(13)$ pix(pc). Filled red circles mark the five cases reported in the top
images, $R_e = 3, 2, 1, 0.5$ and 0.1 pix. The green shaded area on the bottom marks the typical $R_e$ of young massive clusters
observed locally.}
\label{galfit_zoom}
\end{figure}

\subsection{The ultraviolet morphology of D1} 
\label{morph}
\citet{vanz17b} modelled the morphology of D1 using {\tt Galfit} \citep{peng02,peng10}. 
An approximate solution with Sersic index 3.0, $R_e\simeq 140$ pc ($\simeq 8$ pixel, along the tangential direction), 
q (=b/a) $\simeq 0.2$, and a PA of $-28.5$ degrees was found. However, we also noticed a prominent and nucleated core
suggesting that a much compact emitting region is present.
Subsequently, \citet{bouwens18} made use of the HFF observations
to study extremely small objects with a scale of a few ten parsec. D1 was part of their sample, for which they estimate
an effective radius of $38_{-14}^{+21}$ pc (corresponding roughly to 3 pixel along maximum magnification). 
Here we re-analysed in detail the morphology of D1. 

\subsubsection{Empirical half light radius}

We estimated the half-light
radius in the F105W band that is the bluest band (with the narrowest PSF among the near-infrared ones)
probing the stellar continuum redward the \lya\ emission (see Figure~\ref{re}). 
While along the radial direction the profile is consistent with the PSF (FWHM=$0.18''$, or 6 pix), the tangential
profile shows a resolved structure, with a prominent peak containing a large fraction of the UV light.
In particular the observed (one-dimensional) 50\% of the light is enclosed within $\simeq 9$  pix,
suggesting a radius of $\simeq 4.5$ pixel (not PSF-corrected).
If corrected for the PSF-broadening (one-dimensional) the empirical half light radius is 3.4 pix, that
at the redshift of the source (z=6.143) and  $\mu_{tang}=13.2$,
corresponds to $\simeq 44$ parsec (in agreement with \citealt{bouwens18}).
Looking more into the details, the inner region of D1 shows a circular symmetric shape
despite the  large tangential stretch (see, e.g., contours in Figure~\ref{re}), suggesting
a quite nucleated entity significantly contributing to the UV-light (reported below) on top of a more extended {\it envelope}
or dwarf (dubbed D1). In the following we refer to this compact region as D1(core). 
The same highly nucleated region is also evident in the ACS/F814W band, whose PSF ($0.16''$) is slightly narrower than the
WFC3/F105W. Though the intergalactic medium transmission affects half of the F814W band and depress significantly
the overall signal, the S/N of D1(core) is high enough (S/N $\simeq 6$) to still appreciate its compactness (Figure~\ref{galfit_zoom}) 
and, again, well reproduced with a pure HST PSF. In the next section we perform specific simulations to quantify the size
of D1 and its core.

\subsubsection{{\tt Galfit} modelling}

No satisfactory solution can be obtained from a {\tt Galfit} PSF deconvolution of D1 by adopting a single component
(i.e., a single Sersic index, ellipticity q(=b/a), position angle (PA), and effective radius ($R_e$) parameters), 
mainly due to the steep gradient toward the central region.
This reflects the fact that the core appears spatially unresolved, requiring at least two components.
Indeed, a very good model is
obtained combining a Gaussian extended shape that reproduces the diffuse envelope surrounding the core,
and a superimposed PSF$-$like profile which reproduces the central emission (as described in detail in the next sections).
Figure~\ref{galfit} shows the two-components {\tt Galfit} modelling and residuals after subtracting the model from the observed F105W-band image
(for both D1 and T1 objects). 
In the following we focus on the detailed analysis of the size, and eventually the nature, of the nuclear region of D1.
It is worth noting that D1 offers a unique opportunity to access such a nucleated region down to an unprecedented tiny size
for three reasons: (1) it lies in a strongly gravitationally amplified region of the sky ($\mu_{tang}>10$), 
(2) the emitting core is boosted (in terms of S/N) by the underlying well detected envelope (or dwarf), which also implies 
(3) that the detection of the underlying envelope guarantees the full light of the core is captured.
In the next section the shape of the core is specifically addressed. 

\subsubsection{The core of D1: a  source confined within 13 parsec}
\label{magD1}

Depending on the S/N of object and on the knowledge of the PSF, a sub-pixel solution for $R_e$ 
(after PSF deconvolution) can be typically achieved with {\tt Galfit} \citep[e.g.][]{vanz16b, peng10}, as also explored with
dedicated simulations \citep[e.g.,][]{vanz17b}, especially in relatively simple objects showing circular symmetric shapes
like the present case. The central part of D1 is very well detected in all the WFC3/NIR bands, in particular in the F105W band
with a S/N $\gtrsim50$ calculated within a circular aperture of $0.18''$ diameter. 
We adopted two PSFs in the simulations: 1) one extracted from an extensive and dedicated work by \citet{anderson16},
and 2) a PSF extracted by averaging three non-saturated stars present in the same field of the target. The former method benefits from large
statistics and accurate monitoring of the spatial variation along the CCD, the latter includes the same reduction process also applied
to the target D1. The model PSF from Anderson in the F105W band is slightly narrower (FWHM $ \simeq 0.16''$) than the PSF  extracted from 
the stars in the field (FWHM $ \simeq 0.18''$). Both PSFs are useful to monitor the systematic effects in recovering the 
structural parameters as discussed in the appendix~\ref{add_D1}. In the following we adopt the PSF extracted from the stars
present in the same image.

The same procedure described in \citet{vanz16a, vanz17b} is adopted,
where we run {\tt Galfit} on a grid of key parameters like
$R_e$, magnitude and Sersic index $n$, after fixing the position angle (PA), the
ellipticity (q=b/a) and the coordinates of the core (X, Y). These fixed parameters 
are easily determined a priori, especially for objects like the core of D1: circular symmetric and nearly PSF$-$like
(e.g., by running {\tt Galfit} leaving them free at the first iteration).
At each step (i.e., moving in the grid of the parameter space along $R_e$, $n$ and magnitude, with step 0.1, 0.25 and 0.1, respectively)  
the various statistical indicators (standard deviation, mean, median, min/max values) 
have been calculated in a box of $8\times8$ pixel ($0.24'' \times 0.24''$) centered on D1(core) (see Figure~\ref{galfit_zoom}).
The standard deviation and the median signal within the same box calculated in the ``observed-model'' image (image of residuals) are monitored.
 At a given $n$, the smallest standard deviation is reached at the smallest radii, when the residual signal approaches the mean
 value of the underlying, more extended envelope. This is shown in Figure~\ref{galfit_zoom} in which five snapshots
of the residuals on D1(core) at decreasing radii are included. The core is very well subtracted 
using a model with $n=0.5$ (Gaussian shape), magnitude 28.0 and $R_e$ smaller than 1 pix.
The same figure shows the standard deviation as a function of $R_e$,
in which the monotonically decreasing behaviour without a clear minimum indicates that sub-pixel solutions are preferred. 
It is also worth noting that the case of $R_e = 1$ pix still leaves a positive residual suggesting that sub-pixel $R_e$
better matches the D1(core) (Figure~\ref{galfit_zoom}).
Dedicated simulations on mock images quantitatively support this result and provide an upper limit on $R_e$ at sub-pixel scale
(see appendix~\ref{add_D1}).

In particular, Figure~\ref{galfit_zoom} and the simulations described in the appendix~\ref{add_D1} 
(given the S/N and the relatively simple circular symmetric shape) imply that in principle, 
it is possible to resolve D1(core) down to $R_e = 1$ pix. Conversely, D1(core) is not resolved, however we can provide 
a plausible upper limit lower than 1 pix (noting that the cases with $R_e = 0.75$ pix are also recognised in the simulations, 
though with a less success rate, see Figure~\ref{galfit_zoom} and ~\ref{galfit_MC}, red curve).
These limits (0.75/1.0 pix) corresponds to radii $R_e<10-13$ pc at z=6.143, along the tangential direction
discussed above. A size smaller than 2 pix (26 pc) would be a very conservative choice.

It is worth noting that the 25\% of the UV emission of D1 (the entire dwarf) is confined within such a small size (D1(core)), 
suggesting a remarkably dense star formation rate surface density in that region, as discussed in the next section.

\begin{figure*}
\centering
\includegraphics[width=14.5cm]{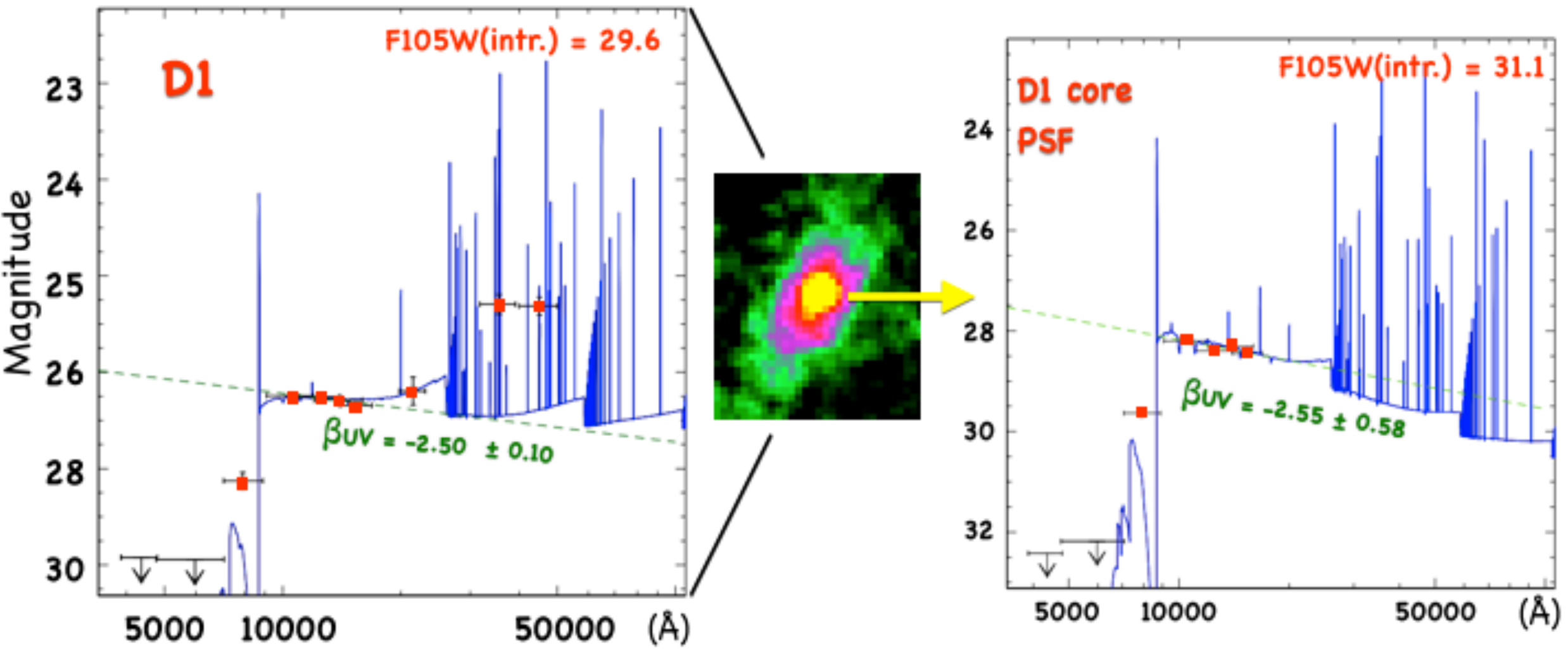} %Fig_SED.pdf}
\caption{The best SED-fit solutions for D1 (left) and for the core of D1 (right) are shown. Only the HST photometry is shown
for the core and no VLT/Ks-band or Spitzer/IRAC magnitudes have been extracted. This comparison shows a first-order consistency 
among the core and the full D1 object, for which very similar ultraviolet slopes ($\beta_{UV}$) are derived. In the middle the F105W image
of D1 and its core (highlighted in yellow) are shown. }
\label{BESTFIT}
\end{figure*}

\subsection{Physical properties of D1}

\begin{figure}
\centering
\includegraphics[width=8.5cm]{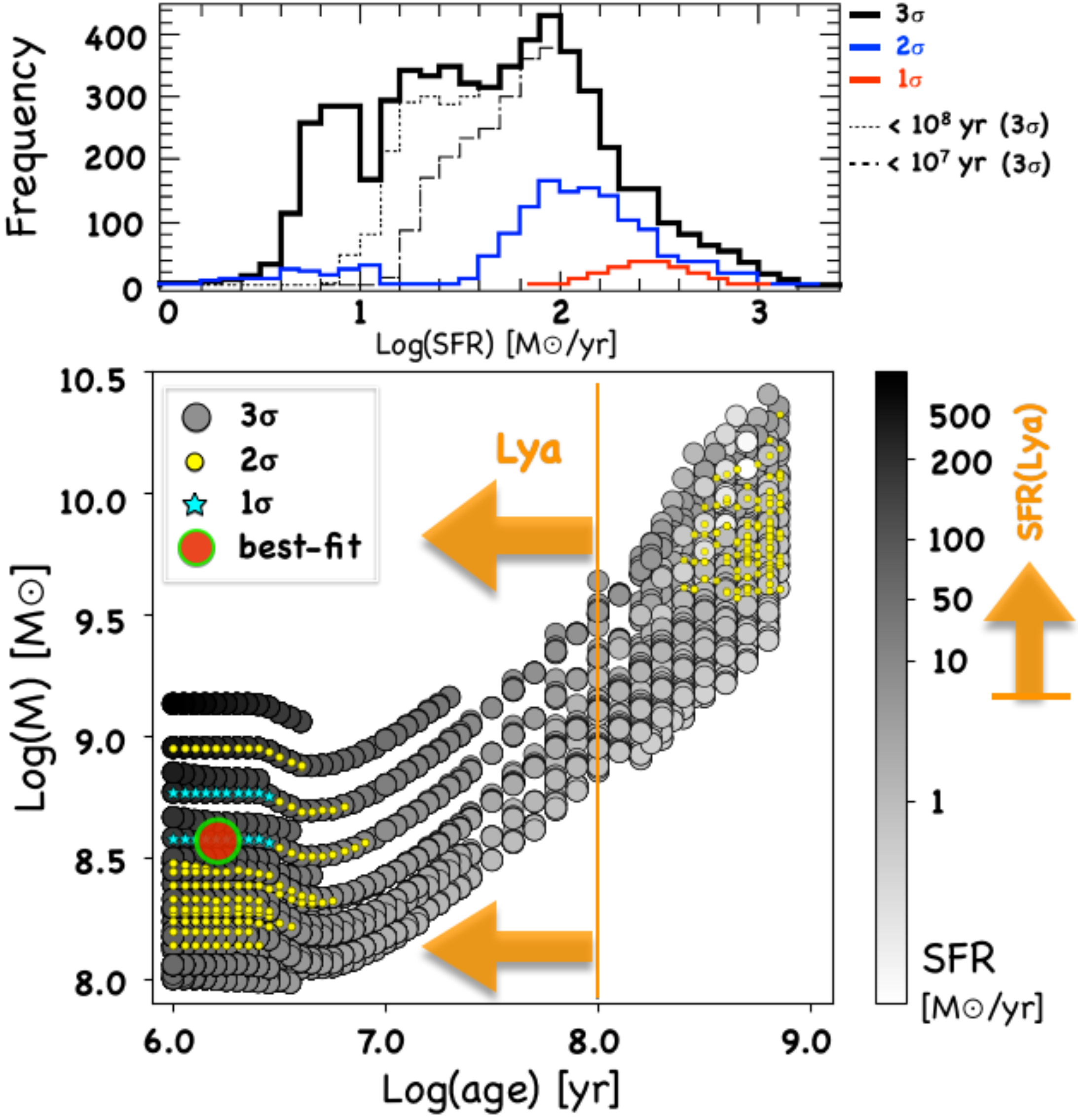} %FIG_degene.pdf} 
\caption{{\it Bottom:} The 3-sigma solutions (99.7\%) derived from the SED-fitting of D1 as a function of the stellar mass, 
age and star-formation rate (grey$-$coded circles). The $2\sigma$ and $1\sigma$ solutions are shown with yellow dots and cyan stars, respectively. 
Solutions with ages younger than 100 Myr and favoured by the
\lya\ equivalent width are indicated with the orange arrows, along with the minimum SFR inferred from the \lya\ luminosity. 
The filled red circle marks the position of the best$-$fit solution, 
for which the best-model is shown in Figure~\ref{BESTFIT}. {\it Top:} The distribution of the star formation rates calculated within
$(3,2,1)\sigma$ is shown for D1. Distributions at 3$\sigma$ with ages younger than 100 Myr and 10 Myr are also shown
with long-dashed and dotted lines. The distribution at 3$\sigma$ with age younger than 100 Myr has been used in the MC calculations
for the estimate of the $\Sigma_{SFR}$, see Sect.~\ref{superdense}.}
\label{degene}
\end{figure}

SED-fitting of D1, based on the Astrodeep photometry
\citep{merlin16} and using nebular prescription \citep{castellano16b} coupled to \citet[][]{BC03} models,  was presented in \citet{vanz17b}
and is shown in the left panel of Figure~\ref{BESTFIT}.
Here we briefly summarise the results, extend the analysis on the degeneracies among the most relevant parameters, thus inferring
the basic properties of D1(core). 
Thanks to the amplification due to gravitational lensing, the faint intrinsic magnitude of D1 (29.60) is placed in a bright regime
(magnitude $\simeq 26.5$) with $\Delta m =$ -2.5Log$_{10}(\mu_{tot}) = 3.1$ ($\mu_{tot}=17.4$). Given the depth of the HFF data,
the resulting S/N is larger than 20 in all the HST/WFC3 bands (from Y to H bands).
As discussed in \citet{vanz17b}, the relatively small  photometric error in the VLT/HAWKI Ks-band (S/N$\simeq 3.5$)
leads to non-degenerate solutions
(within 1$\sigma$) among SFR, stellar mass and  age. 
Table~\ref{tab} summarises the best$-$fit values with the 1$\sigma$ and 3$\sigma$ intervals.
The solutions at 1$\sigma$, 2$\sigma$ and 3$\sigma$ are also shown in Figure~\ref{degene}, in which the degeneracy among the stellar mass, age and star formation
rate is evident when relaxing to 3$\sigma$, mainly due to the lack of constraints at optical rest-fame wavelengths. 
Since the distribution of the SFR changes significantly from 68\%(1$\sigma$) to 99.7\%(3$\sigma$)  intervals, 
we conservatively adopt the 3$\sigma$ distribution for the following calculations. In the next section we will provide additional constraints
on the SFR and the age of the system by considering the  \lya\ emission. 

\subsubsection{Additional constraints from \lya\ emission}
\label{lya}
\begin{figure*}
\centering
\includegraphics[width=16.5cm]{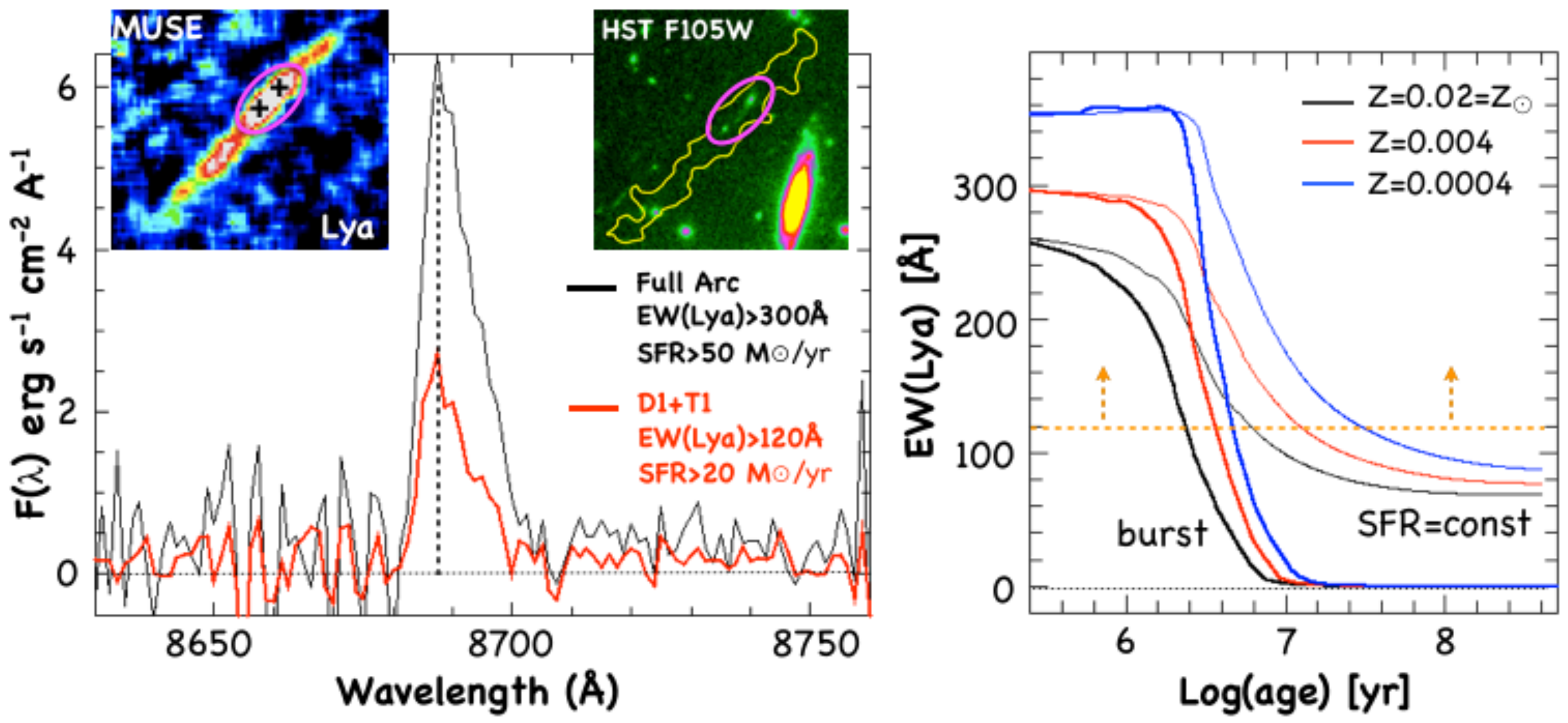} %Fig_EW_LYA.pdf}
\caption{{\it Left}: The \lya\ emission of the system D1T1 extracted from two apertures in the MUSE datacube: {\it local} (magenta ellipse in the top-left inset, red spectrum)
and {\it global} including the full arc (yellow contour in the top-right inset, $>3\sigma$, black spectrum). 
The black crosses in the top-left inset mark the position of D1 and T1 on the \lya\ arc. 
The top-right inset shows the same region in the HST WFC3/F105W band along with the \lya\ contour (yellow line)
and the magenta aperture, from which the \lya\ and UV continuum have been measured to derive the \lya\ 
equivalent width (see text for details). 
{\it Right:} The evolution of the \lya\ equivalent width as a function of time for different star-formation histories and different metallicities, computed
with the \citet[][]{schaerer02} models, assuming a \citet{salpeter55} IMF and upper mass limit of 100 \msun. The horizontal orange dashed line with arrows marks the 
lower limit on the \lya\ equivalent width inferred for the system D1T1 ($>120$\AA).}
\label{EW}
\end{figure*}

Prominent \lya\ emission emerging from the D1T1 complex has been detected in all the three multiple images 
covered by the VLT/MUSE, and follows a well developed arc$-$like shape (Figure~\ref{EW}, see also \citealt{cam17a, vanz17b}).
We calculate the rest-frame equivalent width of the \lya\ line (EW$_{rest}$(\lya)) by integrating the \lya\ flux
and the UV continuum over the same apertures. 
Two estimates of the EW$_{rest}$(\lya) have been derived adopting two apertures: a {\it local} aperture that brackets
the system D1T1 (see the elliptical magenta aperture in Figure~\ref{EW}) and a {\it global} aperture that includes 
the entire \lya\ flux (the yellow $3\sigma$ contour in Figure~\ref{EW}).
The observed line flux for the local(global) aperture is 2.5(6.7)$\times 10^{-17}$ \ergscm\ (with an error smaller than 10\%)
and the magnitude of the continuum at the \lya\ wavelength ($\lambda=8685$\AA)
has been inferred summing up the emission arising from the full system D1T1, $m \simeq 26.0$.
Within the $3\sigma$ contour \lya\ arc, no evident HST counterparts have been identified, besides the D1T1
complex, suggesting that the bulk of the ionising radiation producing the \lya\ arc is generated by this
system.\footnote{Possible additional fainter sources of ionising
radiation may contribute to the total \lya\ flux, however, they would be more than 2.5-3.0 magnitudes fainter than D1 and T1, and
consequently their contribution would be negligible in our analysis. The F105W magnitude of the D1T1 star-forming complex has been
derived from the Astrodeep photometry.}
The magnitude of the continuum 
has also been corrected for the observed UV slope ($\beta = -2.5$, \citealt{vanz17b}). 
The resulting rest-frame EW$_{rest}$(\lya) is $60\pm8$\AA~and $161\pm15$\AA~for the local and global apertures, respectively. 
While these are large values that place complex D1T1  in the realm of  \lya-emitters, the intrinsic EW is plausibly higher than the
observed one, for mainly two reasons: (1) the clear asymmetry of the line profile suggests the bluer part of the line is undergoing radiative transfer effects, being possibly attenuated 
by the intergalactic, circum galactic and/or the interstellar \hi\ gas. 
A factor two attenuation is a conservative assumption at these redshifts \citep[][]{laursen11,debarros17}; 
(2) the best SED fit allows for the presence of low or moderate dust 
attenuation, in the range E(B-V)$_{stellar} \simeq 0.0-0.15$ that would make the observed line flux a lower limit.
 Given the resonant nature of the \lya\ transition that make such a line
 fragile when dust is present and the fact that 
 the dust attenuation would be typically larger for the nebular lines than the stellar continuum \citep[e.g.,][]{calzetti00,hayes11},
 the intrinsic equivalent width of the line is likely higher than observed. The current data prevent us from quantifying the dust attenuation (future observations of the
Balmer lines with JWST will  provide valuable hints on that), therefore we consider the \hi\ attenuation 
only (case 1) and assume no dust absorption, i.e. the inferred EWs are still lower limits due to the possible presence of (even a small amount of) dust.
Therefore plausible lower limits on the equivalent widths are  EW$_{rest}$(\lya)$~>120$\AA~and $>320$\AA~for the local and global
apertures, respectively. 
The presence of such a copious \lya\ emission implies a ionisation field associated to young stellar populations.
Indeed, even in the most conservative case (EW$_{rest}$(\lya) $ > 60(120)$\AA), the comparison with the temporal evolution of the \lya\ equivalent width
extracted from synthesis models suggest an age of the star-forming region(s) younger than 100 Myr, or even younger than 5 Myr in the case of bursty star-formation 
Figure~\ref{EW} shows the EW$_{rest}$(\lya) as a function of the age, metallicity, instantaneous burst and constant star-formation
extracted  from models of \citealt{schaerer02}. The observed \lya\ luminosity also provides
a lower limit on the star-formation rate, assuming the case B recombination applies here \citep[][]{kennicutt98a}. 
The observed \lya\ luminosity for D1T1 is $1.05\pm 0.05 \times 10^{43}$ \ergs\ (derived from the local aperture accounting for the  
factor 2 due to \hi\ attenuation, see the case (1) above) and corresponds to SFR~$ > 20(51)$ \msunyr, in the case of local (or global) aperture (Figure~\ref{EW}). Since we are focusing on the D1 source only, a very conservative lower limit of SFR~$ > 6$ \msunyr\  
has been calculated by integrating the \lya\ flux within a circular aperture of $1''$ diameter centred on D1. 
Figure~\ref{degene} shows the 1$\sigma$, 2$\sigma$ and $3\sigma$ solutions
of the SED-fitting for D1 including the aforementioned constraints inferred from the \lya\ emission (orange arrows).

\subsubsection{A superdense star-forming region hosted by the dwarf galaxy D1}
\label{superdense}

The prominent and nucleated UV emission arising from the core of D1 suggests a particularly high star formation rate surface density
(SFRSD or $\Sigma_{SFR}$ hereafter, [M$_{\odot}$~yr$^{-1}$~kpc$^{-2}$]) which we derive using a Monte Carlo approach that includes the
uncertainties of all relevant parameters.

\noindent $\bullet$ {\it The ultraviolet size.} 
As discussed in Sect.~\ref{magD1}, in the following calculations we consider 1 pix (13 pc) as an upper limit for the effective radius of the nuclear region.
It is worth anticipating, however, that even adopting a more conservative assumption of $R_e = 2$ pix (26 pc), the resulting $\Sigma_{SFR}$ still lies
in the high density regime (see below).

\noindent $\bullet$ {\it The star formation rate.} Figure~\ref{BESTFIT} shows the best SED-fit solutions for D1 and D1(core). In the latter case,
the aperture photometry matching the HST PSF (0.18$''$ diameter)  has been specifically performed. 
Given the lower spatial resolution (respect to HST),
the VLT/Ks and the Spitzer/IRAC bands have not been considered
in the fit. 
The critical condition in which the photometric analysis is performed (very localised region) and the faintness of the object ($m\simeq 28$)
prevent us from deriving solid results from the SED$-$fit procedure directly, that simply mirrors the same degeneracies we see for D1 at 3$\sigma$, 
but here at 1$\sigma$ for D1(core). 
We therefore adopt the SFR derived for D1 (whose SED$-$fit benefits from a much brighter photometry) 
and rescale it accordingly to the flux density ratio in the ultraviolet. 
Specifically, both objects show a fully consistent spectral shape (Figure~\ref{BESTFIT}), as steep as $\beta \simeq -2.5$ 
($-2.50\pm0.10$ and $-2.55 \pm 0.58$ for D1 and D1(core), respectively). 
Given this photometric similarity, the co-spatiality and the \lya\ emission suggesting a relatively short age of the burst,
It is reasonable to assume that they shared a common SFH; in this case, a good proxy
for the SFR of
the core can be obtained by rescaling the SFR of D1 by the measured ultraviolet luminosity density ratio among the two, % in the ultraviolet 
i.e., we adopt proportionality among the ultraviolet luminosity and the SFR \citep[][]{kennicutt98a},
such that L$_{1500}$(core)/L$_{1500}$(D1) $\simeq$ SFR(core)/SFR(D1) $\simeq$ 0.25, 
L$_{1500}$ is derived from the F105W$-$band  on the basis of the morphological analysis discussed above. We assume the uncertainty of the
flux ratio follows a Gaussian distribution with $\sigma = 0.04$, given by the flux error propagation (used
in the MC calculation). We note that the SFR inferred from the SED$-$fitting directly performed on D1(core) spans the 68\% interval 
of $1-40$ \msunyr\ (i.e., 0.06$-$2, \msunyr\ intrinsic, see Appendix~\ref{degene_core}), similar to what obtained by rescaling the global fit of 
D1 as mentioned above.
The stellar mass inferred for D1(core) is $1.5\times 10^{7}$ \msun, i.e., $\simeq 0.86 \times 10^{6}$ \msun\ intrinsic (see Table~\ref{tab}, 
with the usual caveats related to the limited spectral coverage, see Sect.~\ref{future}). Therefore, the stellar mass of D1(core) is
$\lesssim 10^{6}$ \msun\ (Appendix~\ref{degene_core}).

\begin{figure*}
\centering
\includegraphics[width=16.5cm]{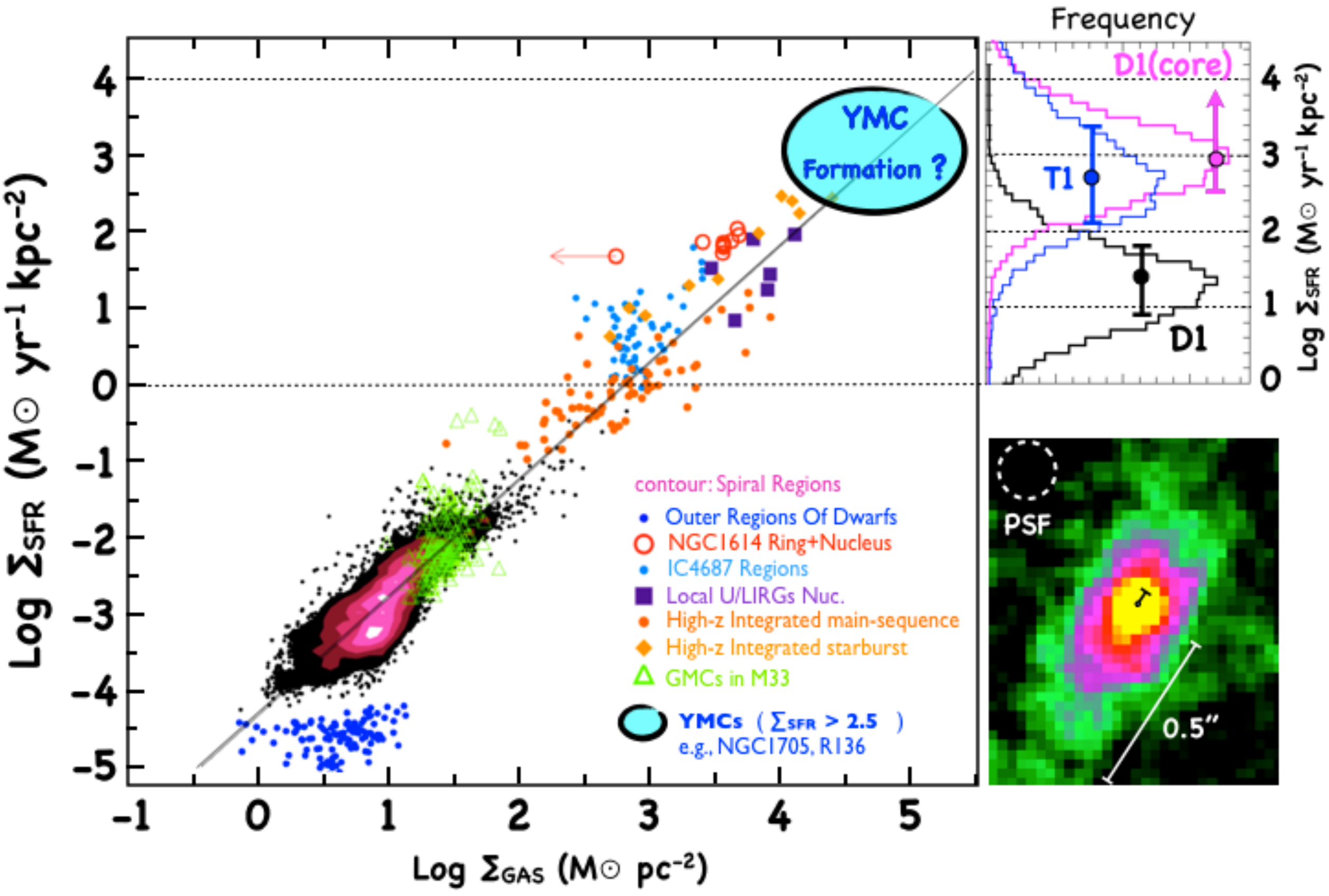} %Fig_SK2.pdf}
\caption{Kennicutt-Schmidt (KS) law from various estimates in the literature  (adapted from \citealt{shi18}).
The SFR surface density distributions for D1, D1(core) and T1 are shown in the top-right panel with black, magenta and blue lines, respectively (see text for details). The filled circles with 68\% central interval mark the medians of the corresponding distributions and are shifted in the X-direction for clarity. Note the magenta point corresponds to a lower limit.
In the bottom-right panel the HST/F105W image of D1 is shown, in which the segments indicate the 26 pc physical size for the core (black line)
and $\sim 200$ pc for D1 (white line). The dashed circle shows the PSF size in the same F105W band ($0.18''$ diameter).}
\label{SK}
\end{figure*}

From three key quantities, i.e., magnification, morphology and the SFR, we derive the $\Sigma_{SFR}$ of the two objects, D1 and D1(core).
The size of D1 has been inferred from the F105W band and corresponds to 
$17 \pm 3$ pix (corresponding to $0.5''$ observed, or $220 \pm 38$ pc in the source plane along the tangential direction).
The size of the core is spatially unresolved with an effective radius less than 13 pc in the source plane and along
the tangential direction. 
The SFR distribution within the 3$\sigma$ interval has been considered after selecting those solutions associated with an
age younger than 100 Myr, as inferred from the \lya\ equivalent width (see Figures~\ref{degene} and~\ref{EW}).\footnote{The results do not change
significantly if we include all the possible SFRs and impose a limit on the age equal to the age of the Universe at z=6.1.}
The SFRSD has been calculated by extracting 10000 values for the tangential magnification $\mu_{tang}$, ultraviolet sizes
and the SFRs, accordingly with best estimates/limits and uncertainties. In particular,  $\mu_{tang}$ is assumed to follow
a Gaussian distribution with mean 13.2 and $\sigma=4.0$ (see Sect.~\ref{lensmodel}). The size of D1 is 
drawn from a Gaussian distribution with mean 17 pixels and $\sigma=3.0$ pixels (see 2$\sigma$ contour shown in Figure~\ref{re}), while in the case of D1(core)
the effective radius of 1(2) pix (or 13(26) pc) is assumed as an upper limit for the size (the 2 pixel as a very conservative assumption).
The SFR has been randomly extracted from 3$\sigma$ distributions
resulting from the
SED fitting as discussed above. While the magnification and the sizes are robustly estimated, 
the SFR is the most uncertain and degenerate
parameter (with age, stellar mass and metallicity), for this reason we relax the interval within which the SFR is drawn, 
thus including also the lower tail of SFRs and less dense solution (see 1, 2 and 3$\sigma$ histograms in Figure~\ref{degene}).
The  same Monte Carlo approach was used to compute the $\Sigma_{SFR}$ of T1,
part of the same star-forming complex.
The results are shown in Figure~\ref{SK}, in which the $\Sigma_{SFR}$ of T1,  D1 and D1(core) are reported in the context of the
Kennicutt-Schmidt (KS) law \citep[][]{kennicutt98b}, noting that currently no
information is available for what concerns the gas surface densities
(an approved ALMA program is ongoing and includes the D1T1 system, P.I. Calura).

While D1 shows a moderate SFRSD, i.e. Log$_{10}(\Sigma_{SFR})_{D1}=1.39^{+0.55}_{-0.56}$, 
the same quantity for D1(core) and T1 are quite large, 
Log$_{10}(\Sigma_{SFR})_{core}>2.5$ and $2.7^{+0.5}_{-0.4}$, respectively.
It is worth noting that $\Sigma_{SFR}$ for D1 and T1 might represent an upper limit {\it if} the true sizes are underestimated, 
whereas $\Sigma_{SFR}$ of D1(core) should be regarded as a lower limit, as this object is spatially unresolved and
well captured over the underlying more diffuse stellar continuum (see Sect.~\ref{morph}).  
In particular, the lower limit derived for the core is 2.9 in the case of $R_e<1$ pix (13 pc), and 2.5 if relaxed to the conservative value of $R_e<2$ pix (26 pc).
We recall that the above values have been calculated selecting the solutions of the SED$-$fit with ages younger than 100 Myr 
(as \lya\ properties suggest, see Sect.~\ref{lya} and Figure~\ref{degene}, top panel), however, even including older ages (corresponding to
lower SFR) the result does not change significantly.

\section{Simulating strongly lensed local YMCs at z=6}

We have assessed the reliability of the above analysis by performing end-to-end image simulations
with the software {\tt SKYLENS} \citep{Meneghetti08,Meneghetti10,plazas18} and following the same approach
described in appendix A of \citet{vanz17b}. This code can be used to
simulate observations with different instrumentation (e.g., HST, JWST, ELT), including the lensing effects
produced by matter distributions along the line of sight to distant sources. 
Here we consider the compact blue galaxy BCD~NGC~1705 as a local proxy for D1, 
and place it in the source plane at z=6.143 at the same position of the source that generates D1,
and then lensed on the sky plane using the same model adopted in this work. 
NGC~1705 contains a relatively massive and young
super star cluster  (SSC) of mass $7.15\times 10^{5}$ \msun, with an age of 15 Myr and $R_e = 4$ pc (measured in the optical F555W band).
About ten more lower mass star clusters ($10^{4} - 10^{5}$ \msun) are present with typically older ages spanning the range 
($>10-1000$ Myr, \citealt{annibali09}). The absolute magnitudes of NGC~1705 galaxy
and the SSC are M$_{UV} \simeq -17.3$ \citep{rifatto95} and $-15.2$ (derived from HST/UV observation of the LEGUS survey, \citealt{calzetti15}),  
with a distance modulus of (m$-$M)=28.54 \citep{tosi01}.
These magnitudes are referred to $\lambda \simeq 2000$\AA, close to the rest-frame wavelength observed in the 
F105W band at z=6.143, $\lambda \simeq1500$ \AA~(we do not apply any correction associated to the spectral slope). 
The estimated absolute magnitudes of D1 and D1(core) are $-17.1$ and $-15.6$, therefore quite close to the UV luminosities 
of NGC~1705 and its SSC.

The bluest band observed in the LEGUS survey (WFC3/F275W) provides the image that we used as a model in our simulation,
in which each pixel corresponds to 1~pc ($0.0396''$ at 5.1 Mpc, \citealt{tosi01}).
Figure~\ref{LENS} (left panel) shows the F275W image of NGC~1705 and a zoomed region of the SSC,
in which the SSC dominates the UV emission. 

We simulated HST observations by adding the modelled lensed dwarf 
to the F105W HFF image 
(rescaled to the magnitude of D1 and reproducing a  S/N consistent 
with the observed one), in four positions near the system D1T1
to facilitate a direct comparison with the real object (see Figure~\ref{LENS}). NGC~1705 is marginally recovered and slightly elongated along
the tangential direction (as expected). A prominent and nucleated emission is evident and corresponds to the position of the SSC.
We performed the same {\tt Galfit} fitting as applied for D1 on these four mock NGC~1705 images and find a satisfactory solution when
the PSF was subtracted (as for D1, see top-right panel of Figure~\ref{LENS}). In practice, similarly to D1, the core of NGC~1705 is not resolved
and an upper limit of $R_e = 13$ pc can be associated (in this case we know the SSC has a radius of 4 pc). 
It is clear from this test that the nucleated region of D1 appears consistent with a spatially unresolved super star cluster, as it
emerges from NGC~1705. Another factor that limits the possibility to 
detect and/or spatially resolve single star clusters under such conditions is the large
differential magnification along radial and tangential directions: two close SSCs aligned along radial direction cannot be
distinguished, while along the tangential direction the current resolution does not allow us to probe single star clusters with radii smaller than 15~pc (at least in this specific case in which $\mu_{tang} \simeq 13$). As discussed in Sect~\ref{future}, a sizeable sample of candidate star clusters observed at higher spatial resolution will alleviate these limitations.

\subsection{An E$-$ELT preview}
\begin{figure*}
\centering
\includegraphics[width=16.5cm]{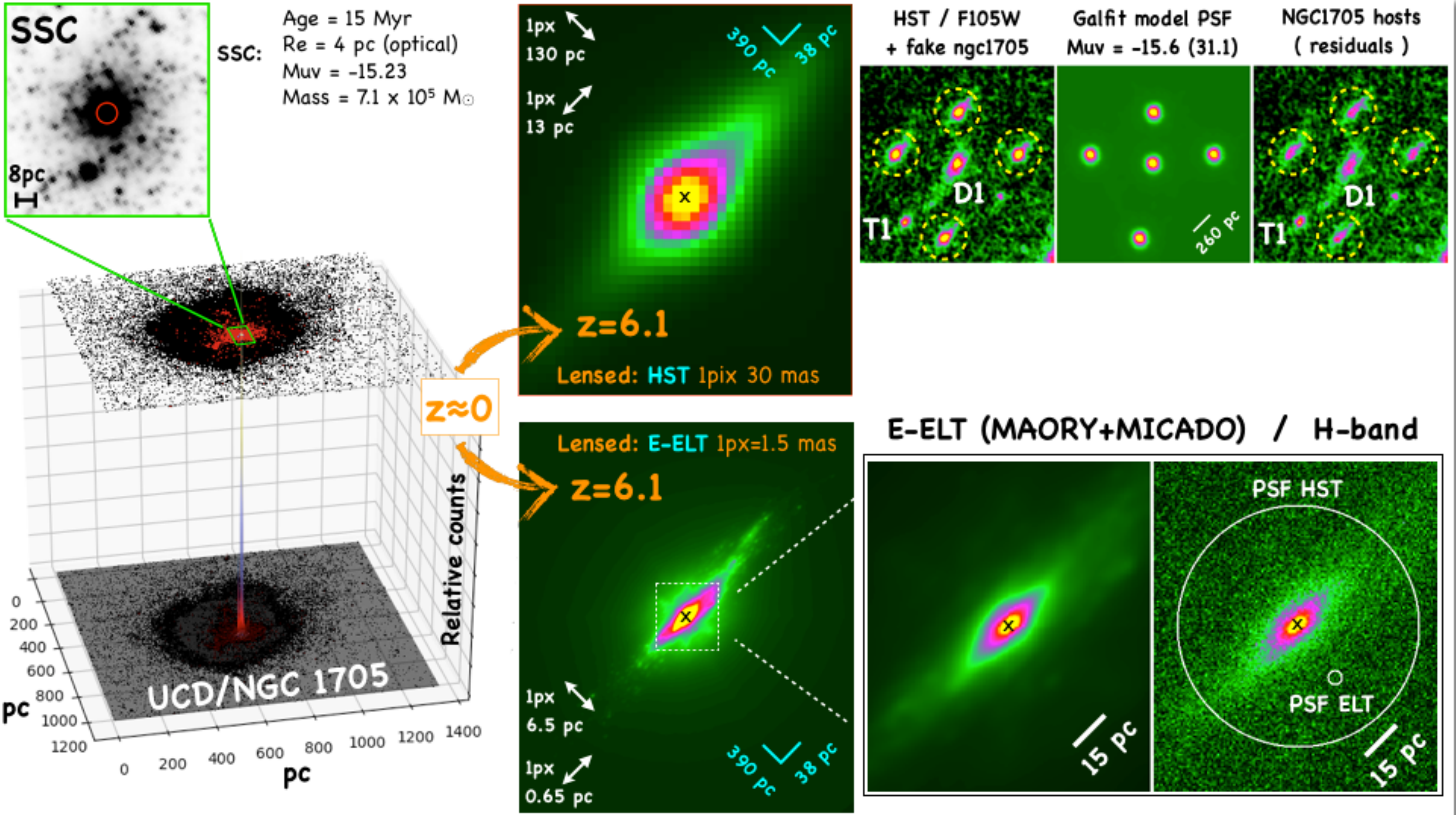} %Fig_SIMUL_LENS4.pdf}
\caption{{\tt SKYLENS} Simulation of the ultra compact dwarf galaxy NGC~1705 hosting its SSC.  In the left panels the WFC3/F275W image
of the galaxy is shown in 3D and 2D with highlighted the prominent UV emission of the SSC (top-left inset). The main properties of the
SSC are also reported. In the top-right panels (from left to right): the modelled noiseless dwarf at the HST resolution lensed at $z=6.143$ in the F105W band 
(the pixel scales are indicated along radial/tangential directions); the same model added to the F105W image in four positions
(dashed yellow circles); the {\tt Galfit} models of the core of NGC~1705 (see text for details); the subtracted PSF models of the previous two images,
showing the unresolved core of the dwarf dominated by the SSC (the position of the SCC is  marked with a black cross in the modelled image).
In the bottom panels, the same simulation is shown adopting the MAORY+MICADO PSF in the MCAO narrow filed mode. The bottom-right panels show the 
zoomed region in which the physical scale and the two PSFs (HST and MICADO+MAORY) are indicated.}
\label{LENS}
\end{figure*}

A significant increase of the spatial resolution will be possible in the future by means 
of extremely large telescopes. 
Figure~\ref{LENS} shows a simulation  of the same lensed dwarf galaxy NGC~1705 
performed considering the 40-m E$-$ELT.  We specifically consider the expected PSF in the H-band of the MICADO camera (Multi-AO Imaging Camera for Deep Observations)
coupled with the MAORY module (Multi-conjugate Adaptive Optics RelaY) adopting the MCAO (Multi-Conjugate Adaptive Optics)
and narrow field mode  ($0.0015''$/pix and FWHM of $\simeq 10$ mas).\footnote{The nominal performances are reported at the following 
link: {\it http://wwwmaory.oabo.inaf.it/index.php/science-pub/}} 
The H and F275W bands probe very similar rest-frame wavelengths,  $\lambda = 2240$\AA\ and $\lambda \sim 2700$\AA.
As shown in Figure~\ref{LENS}, the pixel scale/resolution corresponds to 6.5/40 pc (radial)  and 0.65/4 pc (tangential) in the
specific case of the strongly lensed D1. 
The E$-$ELT PSF ($0.01''$), 18 times smaller than the one of HST in the H$-$band,
and the much larger collecting area lead to a dramatic increase of morphological details.
The noise in the simulation is generated from a Poissonian distribution following the expected performances of the telescope and the MICADO+MAORY instruments.
In particular, a S/N $\simeq 50$ is expected for a point-like object of H=25.6 Vega ($\simeq 27$ AB) and 3 h integration time, within an aperture
of $10 \times 10$ mas. From Sect.~\ref{magD1} the inferred magnitude of D1(core) is $\simeq 28$ (AB) and with the addition of the underlying 
dwarf (D1) the total observed magnitude is 27.25 (AB), or 25.85 Vega. Along the radial direction the expected profile is PSF$-$dominated
($\mu_{rad} \sim 1$), while along tangential direction the resolution is sufficient to resolve NGC~1705 SSC$-$like objects, 
though they will still appear nucleated, as the $R_e$ of the SSC and the resolution element, 10 mas, are similar ($\simeq$ 4 pc).
We therefore expect a S/N slightly lower than 50. 
Although the performances of MICADO, MAORY and the telescope are still under definition,
it is reasonable to expect a S/N for D1(core) in the range 30$<$S/N$<$70 with a few hours integration time,
sufficient to measure the real size of the star cluster.
Figure~\ref{LENS} shows that, depending on the local magnification, a SSC at 
$z\sim6$ will likely be resolved along the tangential direction, as the effective radius
will be sampled with a resolution element of 4 pc (10 mas).
A proper PSF$-$deconvolution (as performed in this work)
should allow to spatially resolve the light profile of the star clusters ($R_e \simeq 4$ pc), in which 1 pix corresponds to $\simeq$ 0.65 pc.  
Possible fainter unresolved substructures will also emerge, 
allowing a proper photometric and spectroscopic analysis of the SSC (i.e. with
significantly reduced confusion).

\section{Discussion}

\subsection{A possible young massive star cluster hosted by D1}

Star clusters are cradles of star formation and grow within giant molecular clouds (GCM)-large collections of
turbulent molecular gas and dust, with masses of $10^{4-7}$ \msun\ and with typical sizes of 10-200 pc.
Previous studies have shown that in the local Universe the fraction of star formation occurring in bound star clusters $-$ usually referred to as the cluster formation efficiency $\Gamma$ \citep{bastian08} $-$
increases with the star-formation rate surface density of the star-forming complexes or galaxies hosting such clusters. 
This emerges from observations of a sample of nearby star-forming galaxies \citep[e.g.][]{messa18,adamo11} and reproduced
in a theoretical framework in which stellar clusters arise naturally at the highest density end of the hierarchy of the interstellar
medium \citep{kruijssen12,li17}.
In particular, $\Gamma$ increases to values higher than 50\% when the SFRSD is Log$_{10}(\Sigma_{SFR}) > 1$, eventually flattening to 
$>90$\% if Log$_{10}(\Sigma_{SFR}) > 2$, a regime in which the density of the gas is so high that nearly only bound structures form \citep{AdamoNate18}.
The $\Gamma-\Sigma_{SFR}$ relation, which reflects the more fundamental $\Gamma-\Sigma_{GAS}$ relation, shows how the galactic and/or the 
star-forming complex environment affects the clustering properties of the star-formation process.

The system D1T1 presented in this work is part of a possible larger structure counting a dozen of individual sources presumably distributed at
$z\sim6$, distributed on a relatively small volume (several tens kpc), and that will be better defined with the ongoing deep MUSE observations (Vanzella et al. in preparation).
If we focus on the system D1T1 and interpret it as a star-forming complex with a size of about 800 pc across (fully including T1 and D1,  Figure~\ref{zoom}), adopting 
the best SFRs estimates reported in Table~\ref{tab}, then the global Log$_{10}(\Sigma_{SFR})_{D1T1} \sim 1.3$ would imply a relatively large
cluster formation efficiency, $\Gamma > 40\%$ (if the relation observed in the local Universe is valid also at high redshift, \citealt{messa18}).
Such a relatively high value is also expected at high redshift \citep[][]{kruijssen12}.
Moreover, the possible ongoing interaction (or merging) between the systems T1 and D1, connected by an elongated structure which looks
like a stellar stream, suggests the presence of young massive clusters, as it has been observed locally in merging galaxies showing also a systematically higher truncation
mass (or upper mass limit) in the initial cluster mass function \citep[e.g., as in the {\it Antennae} galaxies,][]{port10}.

Therefore, putting together the two arguments (high $\Gamma$ and a possible high truncation mass of the initial star cluster mass function in merging systems), 
it would not be surprising that several compact and dense knots, including
the core of D1, T1, and UT1, have been identified within the complex we are investigating and might be the manifestation of a high cluster formation
efficiency (see also UT2 and UT3 knots indicated in Figure~\ref{zoom}). 
The identification of a single gravitationally-bound massive star cluster is the next step and the nucleated emission hosted by D1 and discussed in this work might
support such a possibility, though only future facilities (like JWST and E-ELT) can fully address this issue. However, it is worth noting that the observed stellar mass of D1 ($2\times 10^7$\msun) 
is also consistent with the presence of a single massive cluster (in the present case with a stellar mass of $\sim 10^6$ \msun). 
Indeed, following Eq.~4 and discussion 
in \citet{elme17} \citep[see also,][]{howard18}, 
the total expected mass of a star forming region hosting a single massive cluster with M $=10^6$ \msun\ is $M_{star} \simeq 2\times 10^7$ \msun,
a mass that is fully consistent with what inferred for D1. 
This mass is also compatible for the values expected in some scenarios for GC formation, in which such systems host multiple stellar populations (\citealt{dercole08,calu15,vanz17b} and  Calura et al., in preparation).

Another question we might ask is: what is the evolutionary stage of
the innermost dense forming region?
The inferred $\Sigma_{SFR}$ is extremely high (Log$_{10}(\Sigma_{SFR})>2.5$ or $\simeq 3$) and might suggest it is experiencing the first phases
of star formation in a star$-$cluster$-$like object.

\subsubsection{Comparison with local YMCs: dense star formation}
The inferred $\Sigma_{SFR}$ in the core is consistent with what is expected in the densest star forming young massive clusters (YMCs) observed locally.
A simple estimate of the  $\Sigma_{SFR}$ of young massive clusters hosted in local galaxies (within 10 Mpc distance)
can be derived from the recent release of the catalog of young star clusters observed in the LEGUS survey (P.I. Calzetti, \citealt{calzetti15}),
from which effective radii, stellar masses and ages have been derived for dozens of bound stellar systems and 
in the mass range of $\sim (0.01-1)\times 10^{6}$\msun\ \citep[e.g.,][]{adamo17,ryon17}. 
As an example, the super star cluster hosted by the ultra compact dwarf galaxy NGC1705
shows an effective radius of $R_e = 4$ pc~\footnote{From the LEGUS catalog the concentration index for this cluster, CI, is 1.87, and corresponds
to a 4 pixel effective radius, that at the distance of NGC1705 of 5.1 Mpc  translates to 4 pc, see Figure~4 of \citet{adamo17}.},
a stellar mass of $7.15\times 10^{5}$ \msun\ and an age of 15 Myr.
The $\Sigma_{SFR}$  can be calculated as follows:
($0.5~\times $ M / $\Delta t$) / ($\pi~R_{hm}^2$), where M is the stellar mass of the cluster, the factor 0.5 accounts for the
half mass radius we used in the calculation, $R_{hm}$ (and $R_{hm}=(4/3) R_e$, \citealt{port10}) and $\Delta t$ is the age of the cluster.
The $\Sigma_{SFR}$ calculated for the SSC of NGC1705
is Log$_{10}(\Sigma_{SFR})>2.4$.
Relatively massive young star clusters have been identified in the interacting {\it Antennae} galaxies (NGC 4038/4039),
with stellar masses of a few $10^{6}$ \msun\ and effective radii in the range $R_e \simeq 1-8$ pc.
In particular, the cluster W99-2 reported by \citet[][]{mengel08} \citep[see also][]{port10} with $R_e = 8$ pc, age 6.6 Myr 
and stellar mass $2.63\times 10^6$ \msun\ is among the most massive and largest clusters studied in that merging galaxy, 
having a Log$_{10}(\Sigma_{SFR})>2.75$ (calculated as discussed above).
Clearly the above $\Sigma_{SFR}$ are very conservative lower limits since the bulk of the star formation plausibly occurred on a shorter timescale. 
In the case of the SSC of NGC~1705, if we assume a duration of the burst lower than 5 Myr then
a much larger value is obtained, Log$_{10}(\Sigma_{SFR}) \sim 2.9$, not dissimilar to what we inferred for the D1(core),
and close to the upper edge of the SK$-$law, approaching the maximum Eddington-limited star
formation rate per unit area discussed by \citet[][]{crocker18}. 
Similarly, also W99-2 SSC might have experienced a $\Sigma_{SFR}$ higher than 1000 M$_{\odot}$ yr$^{-1}$ kpc$^{-2}$ 
if the star formation history was confined within the first 3 Myrs (i.e., within the 50\% of its age).

The detection of massive ($\gtrsim 10^6$ \msun) and young ($< 10$ Myr) star cluster populations in {\it late}-stage
mergers such as the {\it Antennae} galaxies (including also Arp 220 and the Mice galaxies NGC 4676 A/B), has been
statistically extended recently with a sample of 22 local LIRGs showing ongoing merging \citep[][]{linden17}. 
In such big merger events, hydrodynamic simulations show that the ISM condition can produce clusters in
the mass range $10^{5.5} <$ M $< 10^{7.5}$ \msun\ \citep{maji17}.
Presumably, in the present case (though at a lower mass regime with respect to LIRGs), the interacting D1T1 {\it early}-stage system 
might contain similar massive star clusters possibly forming during a proto-galaxy phase \citep[e.g.,][]{peebles68}.
The initial star cluster mass function (and cluster formation efficiency) in such
 early conditions (at z=6) is at the moment observationally unknown, however it is possible that interacting systems, such as D1T1, might have  experienced the formation of high-mass star clusters as observed in local mergers. 
Frequent mergers in high-redshift proto-galaxies provide a fertile environment to produce
populations of bound clusters by pushing large gas masses ($10^{5-6}$\msun) collectively to high density, 
at which point it can (rapidly collapse and) turn into stars before stellar feedback can disrupt the clouds \citep[e.g.,][]{kim18}.

\subsubsection{A globular cluster precursor ?}

So far, the search for local analogs of GC precursors has led to inconclusive results \citep{port10,bastian13}, as no convincing evidence of multiple stellar populations has been found in
local YMCs \citep{bastian17}. The search of forming GCs at high redshift is even more challenging, for several reasons. 
First, as a necessary condition, 
YMCs have to be identified and second, the GCP has to be associated in some way.  The first point is now addressable thanks to
a widely improved set of strong lensing models coupled with deep integral field spectroscopy (e.g., VLT/MUSE) and HST multi-band observations
(like the HFFs), such as the case of D1(core) presented in this work.
In addition, the expected occurrence of forming GCs at $z > 3$ is high \citep{vanz17b, renzini17, bouwens18}, and their detectability is feasible nowadays.  
The second point is strongly related to current globular cluster formation theories, 
with key  parameters represented by the original masses and sizes of proto-GCs \citep[see recently,][]{terlevich18}. 

As discussed in the previous sections, it is very plausible that the D1(core) is 
dominated by (or represents itself) a young massive star cluster detected 
in the first few million years after the onset of a burst of star formation. 
D1 extends $\sim 440$ pc and is part of a lager star-forming complex (that includes D1 and T1 of $\sim 800$ pc across) showing
possible interacting components as outlined by the stellar stream connecting D1, T1 and UT1. 
It is worth discussing if D1(core) (the possible SSC with the highest S/N detection we have) and its environment
can present the expected condition of a forming GC.  
Only those clusters that survive the disruption processes and are still dense and
gravitationally bound can likely become the globular clusters we observe today.
Clearly any inference on what D1T1 would appear today is totally model dependent.

First, we notice that the apparent central position of any nucleated star cluster in D1 might be compatible with the scheme
suggested by \citet[]{goodman18} for the formation
of ultracompact dwarf galaxies (UCD), in which one  possible formation path is the tidal threshing of a nucleated elliptical dwarf galaxy,
after massive star clusters (originated in off-centre giant molecular clouds) migrated toward  the centre of the potential well according on
a timescale dictated by dynamical friction \citep{binney87,goodman18}. With a stellar mass of $\sim 2\times 10^{7} \msun$,  an effective radius 
of $\sim 40$pc and age younger than 100 Myr,
D1 might be in the formation phase of an ultracompact dwarf, in particular the M4 and M5 models of \citet[]{goodman18} in terms of 
mass and half mass radius (assuming the half light radius in the UV is not dissimilar then the optical one).
Interestingly, the presence of the companion T1 (at $\sim 500$ pc distance) and a stellar bridge connecting the two objects
(see Figure~\ref{zoom}), may also suggest a possible ongoing interaction, 
mirroring the tidal threshing mentioned above. 

Second, UCDs share many properties with massive globular clusters, such that dwarf$-$globular transition objects might blur the
distinction between compact stellar clusters and dwarfs \citep[e.g.,][]{forbes08, goodman18} and this is the reason
why $-$ in scenarios in which GCs form in dwarfs $-$ high redshift
galaxies at the faint end of the UV luminosity function will inevitably match the same observational conditions as GC precursors.
It is not the scope of this work to establish the link between the presence of young massive clusters in the system
D1T1 and the potential nature of proto-GCs, and perhaps no strong evidence has been found to date, at any redshift.
However, it is fair to say that globular clusters precursors have in good probability already been detected,
but in most cases not recognised, yet.
It is worth noting that some ancient local dwarf galaxies host possible GCs in their cores, suggesting that in some cases,
the star cluster and its environment (or hosting dwarf) survived for the entire cosmic time \citep[e.g.,][]{cusano16,zaritsky16}.

The system reported in the present work, i.e., the super-dense and
compact star forming region of $\lesssim 10^{6}$ \msun\ located in a forming dwarf (D1) undergoing an interaction with a close
companion (T1) surrounded by an extended \lya\ emitting region
represents one of the most promising cradle hosting a GCP
\citep[e.g.,][]{terlevich18, kim18, zick18, goodman18,renzini17,elme12,trenti15,ricotti16}.
The reader is allowed to accomodate our system into their preferred GC formation scenario. 
In any such scenario, the unknown mass of the present-day by-product of D1 (assuming it has survived down to $z=0$ as a gravitationally bound GC) will be determined by whether the entire D1 object or only its most nucleated region (D1(core)) might be regarded as GCP, especially in the light of the mass-budget argument \citep[see, e.g.,][]{renzini15,vanz17b}. 

\subsection{The \lya\ nebulae surrounding the star-forming complex: what's its origin?}

Local YMCs usually host a large population of very massive stars \citep[e.g., R136 in the 30 Doradus,][]{R136}, 
therefore ionising radiation and feedback from young clusters may have important
effects also at large distances  \citep[e.g., hundreds of pc and up to kpc scale,][]{annibali15, smith16}. 
A detailed analysis of the \lya\ emission and spatial variation along the arc will be better
characterised with the AO-assisted MUSE deep lensed field. However,  
the strong \lya\ emission discussed in Sect.~\ref{lya}  (with equivalent width larger than 120\AA~rest-frame) suggests
an intense ionising radiation field consistent with the emission of young stellar populations and a remarkably low opacity
at the $-$ resonant $-$ \lya\ transition, allowing for a large escape fraction of \lya\ photons (EW(\lya) $>$ 100\AA).
Both the presence of dense \hi\ gas and dust would concur to significantly depress the line, whose prominence, 
instead, implies that some feedback is in place, either in the form of outflowing gas that moves \lya\ photons away from the
resonance frequency (and therefore decreasing the amount of scattering and the probability to encounter dust grains) or as already carved
ionised channels that allow \lya\ photons to freely escape and scatter in the circum-galactic medium up to kpc distances.
Such a kpc-scale \lya\ nebula might also be produced by ionising photons that escape from the D1T1 complex
along the same (or similar) transparent routes and induce \lya\ fluorescence \citep[e.g., as it has been observed in a much 
brighter regime at z=4,][]{vanz18}. Only JWST will allow us to observe the same arc at the Balmer \ha\ wavelength, eventually probing
any fluorescing nature (NIRSPEC/IFU observing at 4.7$\mu$m). 
This will address the possible contribution of high-redshift YMCs to the ionising
radiation field far (by several kpc) from regions where the star formation occurs, eventually quantifying the local escaping ionising
radiation and the possible role of GCPs to the ionisation of the intergalactic medium.

\begin{figure}
\centering
\includegraphics[width=8.5cm]{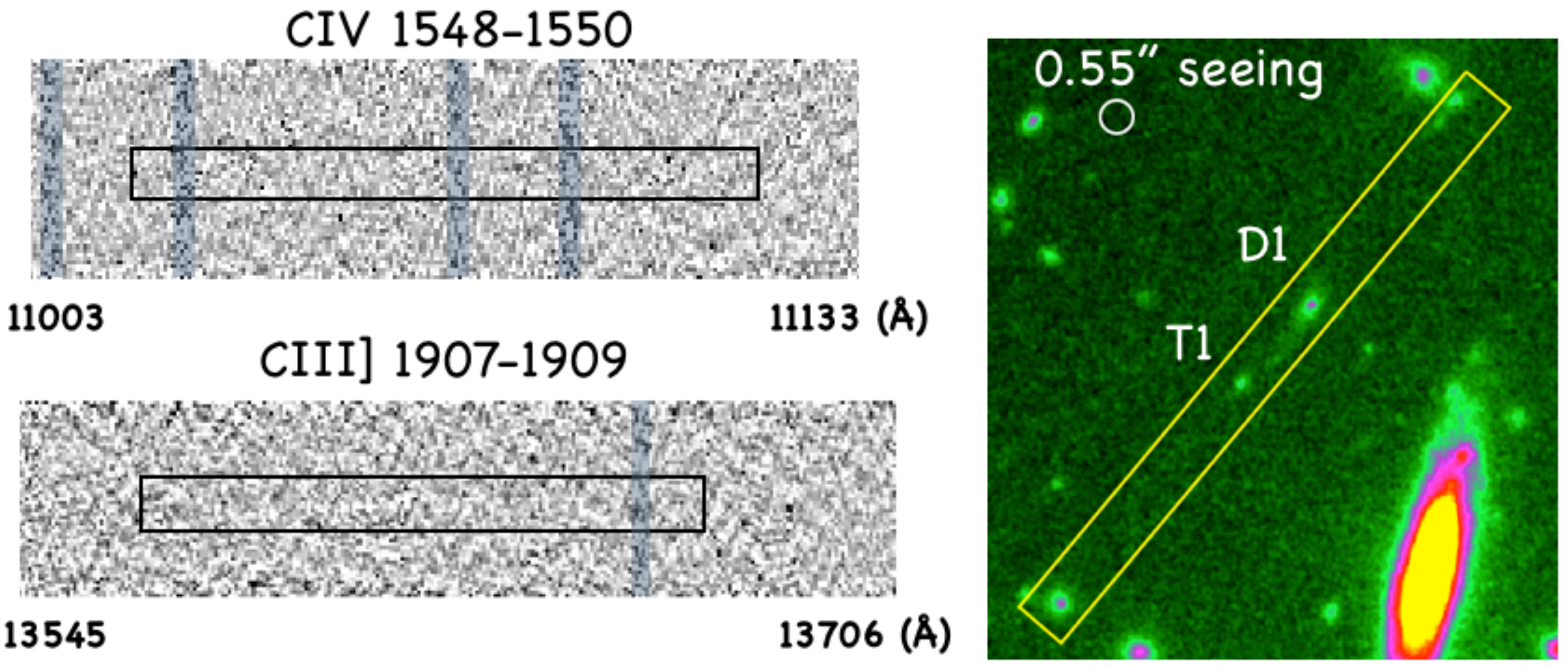} %Fig_xsho.pdf}
\caption{Two hours X-Shooter observations of the system D1T1. On the right side the slit orientation is shown superimposed to the
WFC3/F105W band. The diameter of the white circle corresponds to the average seeing during observations, $0.55''$. A nodding in two positions
of $1.2''$ has been performed. In the left panels the zoomed spectral regions around \civalone\ and \ciiialone\ lines 
are shown. The black boxes mark the positions and the wavelength intervals within $\pm 1000$ \kms\ from the \lya\ redshift (z=6.143).
Vertical gay stripes marks the position of the most prominent sky emission lines.
At the current depth no lines have been detected (see text for more details).}
\label{xsho}
\end{figure}

\subsection{The intermediate mass black hole possibility}

The fact that D1(core) is spatially not resolved leaves room for the possible presence of a faint AGN.
In such conditions, a hosting galaxy with a stellar mass of a few $10^{7}$ \msun\ would imply
an intermediate mass black hole (IMBH) with mass of the order of $ \simeq10^{4}$ \msun\ \citep{kormendy13}. 
Assuming the underlying spectrum for the IMBH is the same as observed in brighter AGNs, the presence of
high-ionisation lines and line ratios could be used to
investigate the nature of the ionising source \citep[e.g.,][]{feltre16,gutkin16}. To this aim, two hours VLT/X-Shooter observations (ID 098.A-0665B, PI: E. Vanzella)
have been spent during September 2017 with optimal seeing conditions, $0.53''$ and $0.57''$ for the two OBs. 
The slit orientation is shown in Figure~\ref{xsho} in which a dithering pattern of $1.2''$ has beed implemented (to avoid superposition among D1 and T1).
Data reduction has been performed adopting the same procedures described in \citet{vanz17c} \citep[see also,][]{vanz16a}. 
These exploratory observations provide no detections of \civ\ and \ciii\ lines down
to $\sim 3\times 10^{-18}$ \ergscm at  the $1\sigma$ level, neither for T1 nor D1, assuming such lines arise at
$z=6.143 \pm 0.025$, i.e., $\pm 1000$ \kms\  from \lya\ emission (Figure~{\ref{xsho}).
While deeper observations are certainly needed to better explore such transitions, including nebular high ionisation lines
of stellar origin (as narrow as a few \kms\ velocity dispersion, \citealt{vanz17c}),
the shallow limits currently available imply \lya\ / \civalone\  $> 8(22)$ for the D1T1 system, 
adopting the \lya\ flux measured in the {\it local}({\it global}) aperture, as discussed
in Sect.~\ref{lya} (flux(\lya)$=2.5(6.7) \times 10^{-17}$ \ergscm), 
possibly excluding any evident AGNs \citep[e.g.,][]{alex13}.
Similarly, no \nv\ has been detected in the MUSE data cube, providing a  \lya\ / \nv\ $> 18$
at the 1$\sigma$ limit. This is a very conservative lower limit if we consider that the \lya\ flux is also a 
lower limit (see Sect.~\ref{lya}). This limit is higher than the typical values reported
for AGNs at moderate ($z\sim 2-3$) and high ($z>5$) redshifts \citep[see discussion in][]{castellano18}.
While the possible presence of an IMBH would be extremely interesting, being such
objects never been observed (especially at high redshift) and representing a current challenge for the theoretical models of BH and structure formation
 \citep[e.g.,][]{reines16,pacucci18}, 
 the current data are consistent with the star-cluster interpretation.

\begin{table}
\footnotesize
\caption{The inferred physical, morphological and lensing properties of D1 and its compact SF-region (core). 
Also the properties of the local dwarf galaxy NGV~1705 are reported.}
\begin{tabular}{l l r} 
\hline
\hline
Quantity  & Best value &  uncertainty\\
\hline
$\mu_{tot}$ (magnif.)       & 17.4   &  $\pm 1 _{stat}$ $\pm 5_{syst}$ \\
$\mu_{tang}$ (magnif.)    & 13.2   &  $\pm 0.5 _{stat}$ $\pm 4_{syst}$ \\
\hline
D1(total) & & \\
\hline
M(stellar) [$\times 10{^8}~M_{\odot}$] 1$\sigma$ &  3.8 $\mu_{tot}^{-1}$ & [$3.7-5.8$] $\mu_{tot}^{-1}$  \\    
M(stellar) [$\times 10{^8}~M_{\odot}$] 3$\sigma$ &                                    & [$1.0-250$] $\mu_{tot}^{-1}$  \\  
Age [Myr]  1$\sigma$          &  1.4  & [$1-3$]        \\    
Age [Myr]  3$\sigma$          &         & [$1-708$]        \\    
SFR [$M_{\odot} yr^{-1}$] 1$\sigma$  &  275 $\mu_{tot}^{-1}$ & [$131-585$] $\mu_{tot}^{-1}$  \\ 
SFR [$M_{\odot} yr^{-1}$] 3$\sigma$  &                                    & [$1-1350$] $\mu_{tot}^{-1}$  \\ 
E(B-V)  1$\sigma$                               & $0.15$           & [$0.15-0.20$] \\   
E(B-V)  3$\sigma$                               &                       & [$0.0-0.30$] \\   
$m$(1500\AA)(intrinsic)  & 29.60 & $\pm 0.2$\\ 
$M_{UV}$(1500) & $-17.13$  &   $\pm 0.2 $\\   
log($\Sigma_{SFR})^\star$ & 1.39 & $[0.80-1.85]$ \\
$R_e$ tang. [pix(pc)]  & 3.4(44)$^{\star \star}$              &  $\pm1.5$($\pm 19$) \\
Half-Size tang. [pix(pc)] & 17(220)$^{\star \star}$            &  $\pm3$($\pm 35$) \\
\hline
D1(core) & & Comment\\
\hline
M(stellar) [$\times 10{^7}~M_{\odot}$] 1$\sigma$ &  $\simeq$ 1.5 $\mu_{tot}^{-1}$ & $-$  \\   
$m$(1500\AA)(intrinsic)                      & 31.10 & $\pm 0.3$\\ 
$M_{UV}$(1500) & $-15.6$  &   $\pm 0.3 $\\   
log($\Sigma_{SFR})^\star$  & $>2.5$& $R_e<26$ pc (2 px)\\
log($\Sigma_{SFR})^\star$ & $>2.9$& $R_e<13$ pc (1 px)\\ 
$R_e$ tang. [pix(pc)]  & $<$ 1.0(13)$^{\star \star}$   &  PSF-shape \\  
\hline
NGC1705 \& SSC & & \\
\hline
$M_{UV}$(2000)(NGC1705) & $-17.3$  &   $\pm 0.1 $\\   
$M_{UV}$(2000)(SSC) & $-15.2$  &   $\pm 0.1 $\\   
log($\Sigma_{SFR})^\star$(SSC) & $>$2.6 \\ 
$R_e$ [pc]  & $4.0$   &   F555W-band\\  
\hline
\hline
\end{tabular}
$^\star$ $\Sigma_{SFR}$ in units of M$_{\odot}$yr$^{-1}$kpc$^{-2}$.\\
$^{\star \star}$ $R_e$[pc] = $R_e$[pix] $\times~0.03'' \times 5660$ pc~/~$\mu_{tang}$; 1pix$=0.03''$; $1''=5660$pc at z=6.14\\
\label{tab}
\end{table}

\section{Summary and conclusions}

We studied an ideal case in which a dwarf galaxy hosting a nucleated ultraviolet emission 
is significantly stretched due to gravitational lensing.
Firm constraints on a star-forming region of unprecedented small size at z=6.1 have been achieved.  
The present constraints pave the way  towards a possible future unambiguous detection of a forming super star cluster
in the first Gyr of the universe. In particular:

\begin{figure}
\centering
\includegraphics[width=8.5cm]{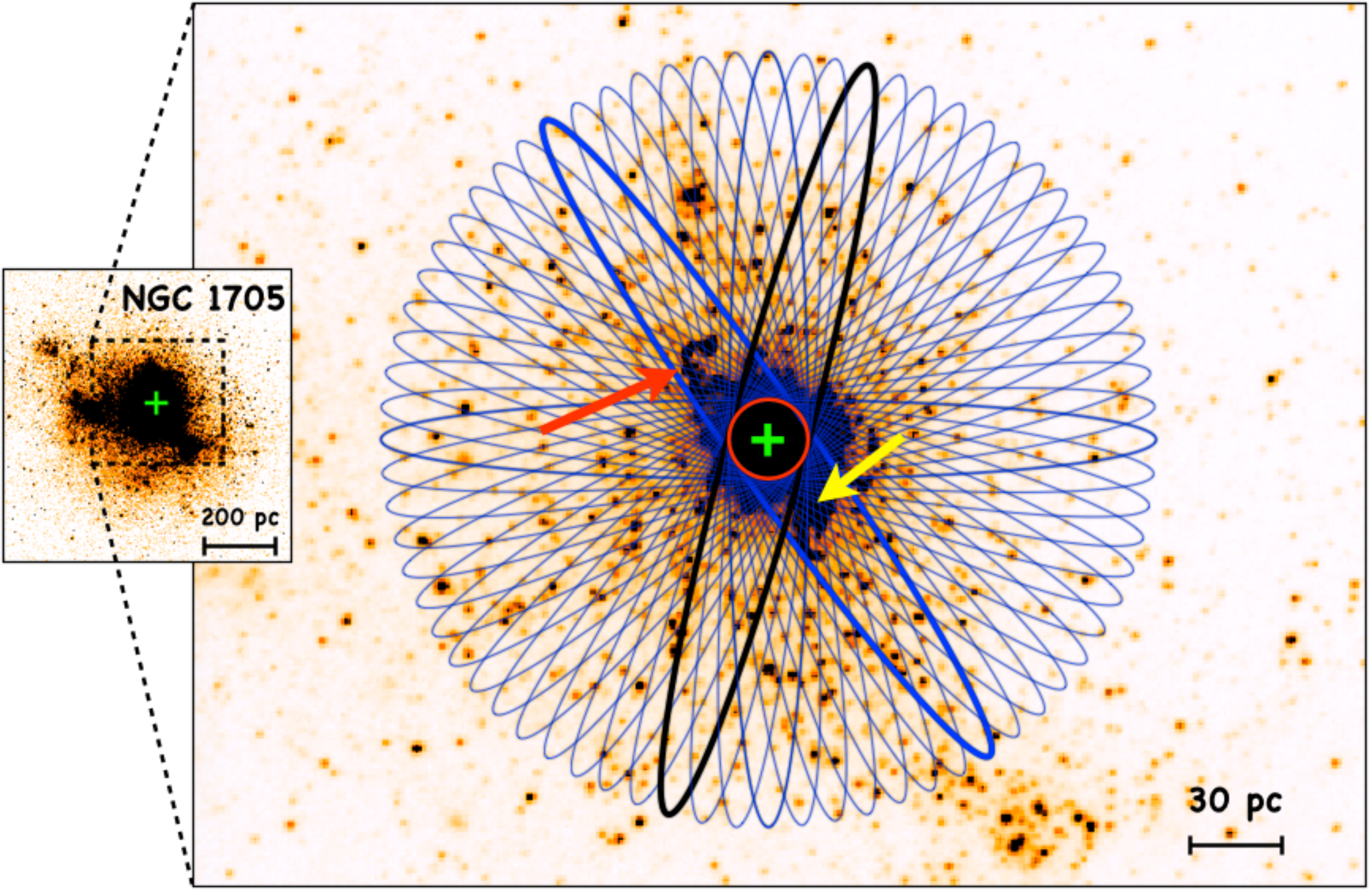} %Fig_xsho.pdf}
\caption{A calculation of light contamination from star-forming complexes surrounding the SSC 
due to lensing stretch ($\mu_{tang}/ \mu_{rad} \simeq 10$). The blue ellipses have semi-minor and semi-major
axes of 13 pc and 130 pc, corresponding to the limits we calculated along tangential and radial directions for the z=6.14 arc,
respectively. The median flux is computed from 36 ellipses placed with different position angles (with step of 5 degrees from each other) 
and all centred on the SSC (marked with a green cross), and subsequently compared to the flux derived from the circular aperture 
of 13 pc radius (red circle), centred at the same position. The maximum/minimum ratio (counts(ellipse)/counts(circle)) happens
at the  blue/black thick ellipse, corresponding to overestimated fluxes by factors 1.52 and 1.30, respectively 
(see the text for more details). The maximum ratio is given by the inclusion in the ellipse of two structures:
a star-forming complex (indicated with a red arrow) and a star cluster (indicated with a yellow arrow, dubbed star cluster ``1" 
by \citealt{annibali09}) both at $\simeq 23$ pc away from the main SSC. An image of NGC~1705 is shown in the leftmost inset, in which the dashed square outline
the zoomed region shown in the  main panel.}
\label{ellipse}
\end{figure}

\begin{itemize}
\item{A superdense and compact star-forming region, D1(core),  with $R_e < 13$ pc 
in a dwarf galaxy (D1, extending up to $\sim 440$ pc) is confirmed at z=6.143, 
which is, in turn, part of a larger and interacting star-forming complex that includes D1 and T1 (extending to $\sim 800$ pc across). 
D1 and T1 are spatially resolved and connected by a stellar stream also containing an ultrafaint star-forming knot
(indicated as UT1, Figure~\ref{zoom}).   
The D1(core) shows a circular symmetric shape fully consistent with the HST PSF despite the large gravitational stretch
well depicted by the giant tangential \lya\ arc.}
\item{Extensive {\tt Galfit} modelling and MC simulations clearly demonstrate the compactness of D1(core) that also accounts for the $(25-30)$\% of the
ultraviolet light of the entire dwarf galaxy D1, implying a very large star-formation rate surface density occurred in such a compact region.
After including realistic uncertainties on the magnification, morphology and the SFR, the star-formation rate surface density is quite large,
$\Sigma_{SFR} > 300$ M$_{\odot}$~yr$^{-1}$~kpc$^{-2}$ in the conservative case, or  $\Sigma_{SFR} > 800$ M$_{\odot}$~yr$^{-1}$~kpc$^{-2}$
as a best estimate.
The comparison of the same expected quantity derived for local young massive clusters  during their formation 
phase ($\Sigma_{SFR} \simeq 1000$ M$_{\odot}$~yr$^{-1}$~kpc$^{-2}$) suggests the D1(core) is fully compatible with being
a fresh super star cluster with a stellar mass of $\lesssim 10^6$ \msun\ and observed just a few Myr after the onset of the star-formation.
The stellar mass of the hosting galaxy D1 of $2 \times 10^7$ \msun\ is also consistent with the presence of such a massive cluster. 
Accordingly to several scenarios, such a system is an ideal candidate globular cluster precursor.}
\end{itemize}

The ultraviolet appearance of D1 and its core also match those of the ultra compact dwarf galaxy
NGC~1705 simulated at z=6.143 and strongly lensed via the {\tt SKYLENS} tool through
our currently best lens model for the HFF cluster MACS~J0416. 
NGC~1705 contains a well known young super star cluster. Both the host galaxy, NGC~1705, and its SSC, show
very similar luminosities and masses as estimated for D1 and D1(core), respectively (see Table~\ref{tab}).

The prominent \lya\ emission of equivalent width larger than 120\AA~rest-frame surrounding the system D1T1, 
extending up to 5 kpc (much larger than the detected stellar continuum of the D1T1 stellar  complex, $\lesssim 1$ kpc) 
and the very blue ultraviolet spectral slope ($\beta = -2.50\pm0.10$) suggest the underlying
 stellar populations are very young ($<10-100$ Myr) and with very little dust attenuation.
\lya\ photons are very sensitive to dust absorption, especially in presene of 
very dense \hi\ gas \citep[e.g.,][]{verhamme06}, as the high $\Sigma_{SFR} $ seems to imply. 
An enhanced \lya\ visibility can be related to ongoing feedback through 
outflows and/or carved ionised channels. However, the real nature of the \lya\ nebula is still unknown: 
it can be the result of pure \lya\ scattering or fluorescence 
induced by escaping ionising radiation, or both.  Fluorescence is also connected to the ionisation power of such tiny sources
in the framework of cosmic reionisation, certainly worth investigating in the future.

\subsection{Caveats and future prospects}
\label{future}

While the investigation of parsec-scale ($\lesssim 20$ pc) star-forming regions at $z=2-6$ represent the 
state-of-the-art analysis, joining deep HST imaging, strong lensing and integral field spectroscopy, there are still caveats that limit the current studies. 
We list in the following the significant improvements on the specific system studied here (D1T1) 
that can be achieved with future facilities. The same considerations are equally applicable to the most
general framework of star-cluster searches at cosmological distances and their influence to the surrounding
medium (feedback, ionisation).

\begin{itemize}
\item{{\it Spatial resolution:} it is unknown what is the distribution of the size ($R_e$) of the most compact SF regions currently 
identified at high-redshift, like the core of D1. A significant leap will be performed with future instrumentation which will provide
spatial resolution down to 20 mas (e.g., VLT/MAVIS in the optical, $\lambda < 1 \mu m$) or 10 mas (E-ELT/MAORY+MICADO in the near-infrared
with MCAO narrow field mode and much larger collecting area).
Such PSFs, in the specific case (of D1) and along the maximum magnification, will probe light profiles down to $4-8$ pc resolution,
eventually resolving the small, possibly gravitationally bound, star clusters. Ground-based high spatial resolution imaging will be
limited to the blue/ultraviolet at $z>5$ (the K-band probing $\lambda < 3800$ \AA\ rest-frame). In the case of D1T1, 
only JWST will provide morphological information at optical rest-frame wavelengths ($\simeq 6000-8000$\AA), in which the size and
the stellar mass can be properly estimated, as well as the stellar mass surface density.}
\item{{\it Spectral coverage:} while at $z\sim 3$ the accessible spectral range from ground-based facilities still covers the rest-frame
optical (and marginally the \ha\ line, though limited to $z<2.5$), at $z>4$ the rest-frame optical (e.g., the Balmer lines \hb, \ha\ and metal lines) is
redshifted into the $> 2.4 \mu m$ range. In the case of D1T1 the \ha\ lies at 4.66 $\mu m$, observable only with JWST. The \ha\ line
is the best SFR indicator, and by means of the 
 NIRSPEC/IFU (integral filed unit onboard JWST) the nature of the \lya\ nebula (scattering vs. fluorescence) can be investigated. 
The access to the optical rest-frame, both with
imaging and spectroscopy, will also definitely improve the estimate of the physical properties of such potential GCPs 
(e.g., stellar mass, age, SFR, dust content and nebular attenuation via the Balmer lines decrement).}
\item{{\it Statistics:} Despite the power of strong lensing and the positive prospects in the improvement of future lensing models,
large distortion of the images (e.g., large values of $\mu_{tang}/ \mu_{rad}$) might hide features potentially missed in a single object. 
Indeed, in the case of D1T1, whatever small source lying along the radial direction within $\sim 130$ pc from D1 
would be spatially unresolved or merged into a single object, i.e. it would be PSF$-$dominated. 
In such a case the inferred luminosity of the star cluster would be overestimated. In particular, 
as a test-case, we inferred the amount of light contamination for NGC~1705 by calculating the flux within
ellipses of 13 pc~$ \times$~130 pc (of semi-minor and semi-major axis) centred on the SSC and placed over 36 position angles
(with a step of 5 degrees), and compared to the flux measured on a circular aperture of 13 pc radius centred on the same 
SSC (see Figure~\ref{ellipse}). 
On average the light measured on the elliptical apertures is overestimated by a median factor $1.36_{-0.06}^{+0.16}$ (68\% interval), 
corresponding to $\sim$ 0.34 magnitudes. Therefore, assuming the analogy among D1 and NGC~1705, this test would imply the
star cluster hosted in the core of D1 would be $\simeq$ 0.3-0.4 magnitudes fainter than what is measured (corresponding to
$M_{UV} \simeq -15.2$). An even worse case would correspond to multiple star clusters aligned along the radial direction and not
spatially resolved. Assuming a pair of (equal-mass) not spatially resolved SSCs, the true luminosity of each one
would be overestimated by a factor 2 (or 0.75 magnitudes). As mentioned above, future observations with high spatial resolution
facilities (like ELT and VLT/MAVIS) will dramatically improve the situation.
This orientation effect, however, can be washed out after averaging over several D1T1$-$like sources and/or candidate star
clusters/ star-forming complexes. 
From the initial analysis presented in this work, the compactness of several individual knots in the z=6.14 system, spanning the interval
13 pc $<R_e<$ 50 pc, already suggest that D1(core), T1, UT1 (including T3 and T4, see Sect.~\ref{lensmodel}) are intrinsically
small objects, unless all of them are elongated along the same radial direction.
The increasing effort with dedicated HST programs focusing on galaxy clusters (like , CLASH, HFFs, RELICS, e.g.,  \citealt{post12,lotz16,coe18}) 
will timely produce the statistical significant
sample for the current facilities, like ALMA, and the forthcoming ELT and JWST. VLT/MUSE will continue to play a crucial role in the
spectroscopic identification of hundreds of multiple faint images in the redshift range $3<z<6.5$, useful for tuning the lens models and
extending the discovery space of tiny star clusters, and eventually the population of GCPs.}

\end{itemize}

Before the advent of JWST and E$-$ELT,  further progress can be achieved by performing deep spectroscopy in the near-infrared
wavelengths and look for nebular high$-$ionization lines. Indeed, such candidate super star cluster(s) may contain significant
amount of massive and hot stars (T$_{eff} > 50$ kK), sufficiently hot to emit photons more energetic than 4~Ryd and produce emission lines
up to the \heii\  (as observed in the case of the young massive cluster R136 in 30 Doradus, \citealt{R136}, and
more recently \citealt{lennon18}).

\section*{Acknowledgments}
We thank the referee for providing constructive comments. 
We thank M. Tosi, L. Hunt, F. Annibali, A. Adamo and F. Cusano for the very 
useful discussions about local dwarf star forming galaxies and young massive star clusters
and P. Ciliegi, M. Bellazzini and E. Diolaiti for useful discussions about the MAORY specifications. 
EV thanks A. Ferrara, N. Gnedin, R. Ellis, H. Katz, D. Sobral and R. Bouwens for stimulating
discussions during the Munich Institute for Astro- and Particle Physics (MIAPP) workshop about the nature of the tiny
 sources presented in this work, and  
 A. Renzini, N. Bastian and E. Dalessandro for sharing their thoughts about the GCPs.
F.C., A.M acknowledge funding from the INAF PRIN-SKA 2017 program 1.05.01.88.04.
MM acknowledges support from the Italian  Ministry of Foreign Affairs and
International Cooperation, Directorate General for Country Promotion.
Based on observations collected at the European Southern Observatory for Astronomical
research in the Southern Hemisphere under ESO programme 098.A-0665(B).

%\appendix
%\section{Additional morphological analysis of D1}

\appendix
\section{Additional morphological analysis of D1}
\label{add_D1}

In this section we run {\tt Galfit} on mock images inserted nearby D1.
1000 fake images have been inserted near D1, each of them resembling the D1(core) in both terms of magnitude and light profile index ($n=0.5$),
but adopting different $R_e$. The simulated images have been inserted avoiding evident sources already present in the F105W image 
(extracted from the Astrodeep catalog, \citealt{castellano16b}). We then performed the simulation in two ways: 
(1) the procedure described in Sect.~\ref{magD1} has been performed for each image by running {\tt Galfit} on a grid of parameters and
deriving the customised standard deviation and (2) by running {\tt Galfit} leaving its internal minimisation procedure and
starting from free parameters.
The aim is to replicate on mock images the same procedures we used for  D1(core) (Sect.~\ref{magD1}).

In case (1) the behaviour of the standard deviation is shown as a function of the $R_e$ for five sets of 1000 images, 
$R_e = 3, 2, 1, 0.75, 0.5$ pixel and using two PSFs. The best solutions (minimum error) occur at $R_e \lesssim 1$ pix.  The sub-pixel cases are marginally recovered and the minimum error disappears as half a pixel is
approached (see Figure~\ref{galfit_MC}). 
When compared to the results we obtained on D1(core) above, this exercise  suggests that $R_e$ is smaller than 1 pixel
(otherwise it would have been recognised).
These tests have been performed using the same PSF in the construction of the mock images and in the recovering
process and as expected   
no systematic errors are present (Figure~\ref{galfit_MC} shows the best solutions that fall exactly on the input radii). 
We then performed the same simulation by using
our PSF (FWHM $=0.18''$) and the Anderson model (FWHM $=0.16''$). 
The effect is shown in Figure~\ref{galfit_MC} (left panel) with red arrows indicating the shifted position
of the minimum standard deviation. Clearly, if a broader/narrower PSF is adopted in recovering
the morphology, the resulting $R_e$ would be underestimated/overestimated. The PSF constructed using the stars present in
the field (and hence including the same reduction process as for the target)  is the closest to the effective PSF in our image 
(FWHM $= 0.18''$, that is also consistent with what calculated by \citealt{merlin16} using the same HFF data).  % suitable in this case.
A similar result is achieved in the case (2), where {\tt Galfit}
recovers the input radius down to $R_e = 1$ pix.
{\tt Galfit} has been run by fixing PA and q (bottom panel of Figure~\ref{MC}) or by fixing PA, q and $n=0.5$ (top panel of Figure~\ref{MC}).
In particular, a solution is reached for 100\% of the 1000 images
in the case of  $R_e = 3, 2$ and 1 pix  (with fixed PA, q and $n=0.5$).
The sub-pixel images ($R_e = 0.5, 0.75$ pix) are partially recovered, though with an increasing
failure as $R_e$ approaches half a pixel (a failure in the fit is provided by an internal warning flag produced by {\tt Galfit}, \citealt{peng10}).

\begin{figure}
\centering
\includegraphics[width=8.5cm]{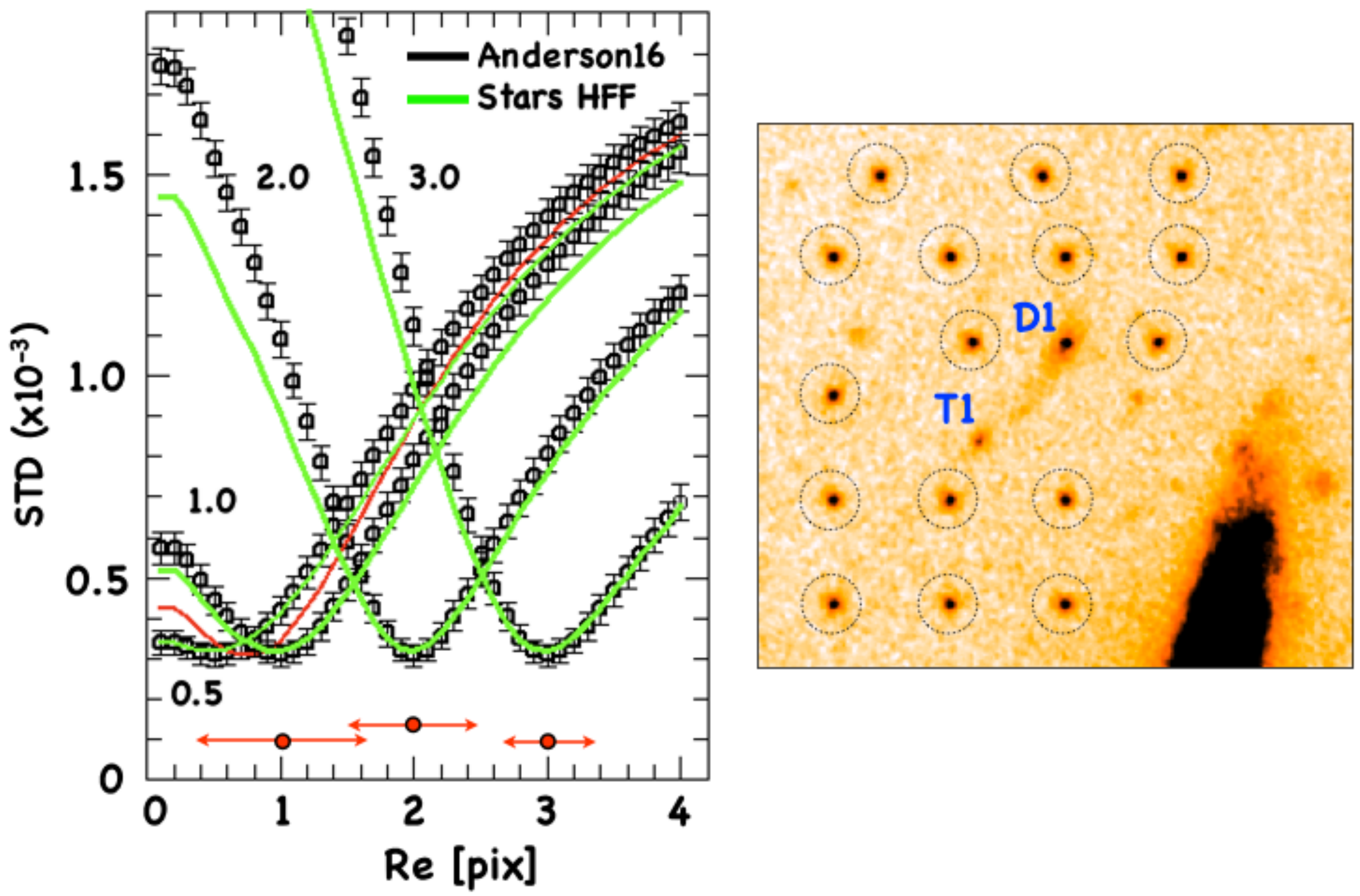} 
\caption{The right panel shows an example of mock images used in the MC simulation (dotted circles). 
On the left side we show 
the standard deviation as a function of $R_e$ (in the interval $0.1-4$ pix with $\Delta R_e=0.1$ pix), from MC simulations of 1000 fake images with different input $R_e = 3, 2, 1$ and 0.5 pix, as done for D1(core) (Figure~\ref{galfit_zoom}).
The black circles with errorbars are the averaged standard deviation and dispersion of 1000 {\tt Galfit} runs, at fixed $n$ and magnitude. The input $R_e$ values are reported near the curves (in pixel) and are
well recovered with the minimum standard deviation in correspondence of the input $R_e$. Sub-pixel radii are marginally (or not)
recovered when half a pixel is approached. Black and green colours represent the same MC performed with two PSFs, 
the model PSF from Anderson and our HFF-based one. 
The horizontal red arrows (with arbitrary position in the Y-axis) show the systematic effect (direction and module) in recovering the input $R_e$ when a different
PSF (wider or narrower) are used in the reconstruction.}
\label{galfit_MC}
\end{figure}

\begin{figure}
\centering
\includegraphics[width=8.5cm]{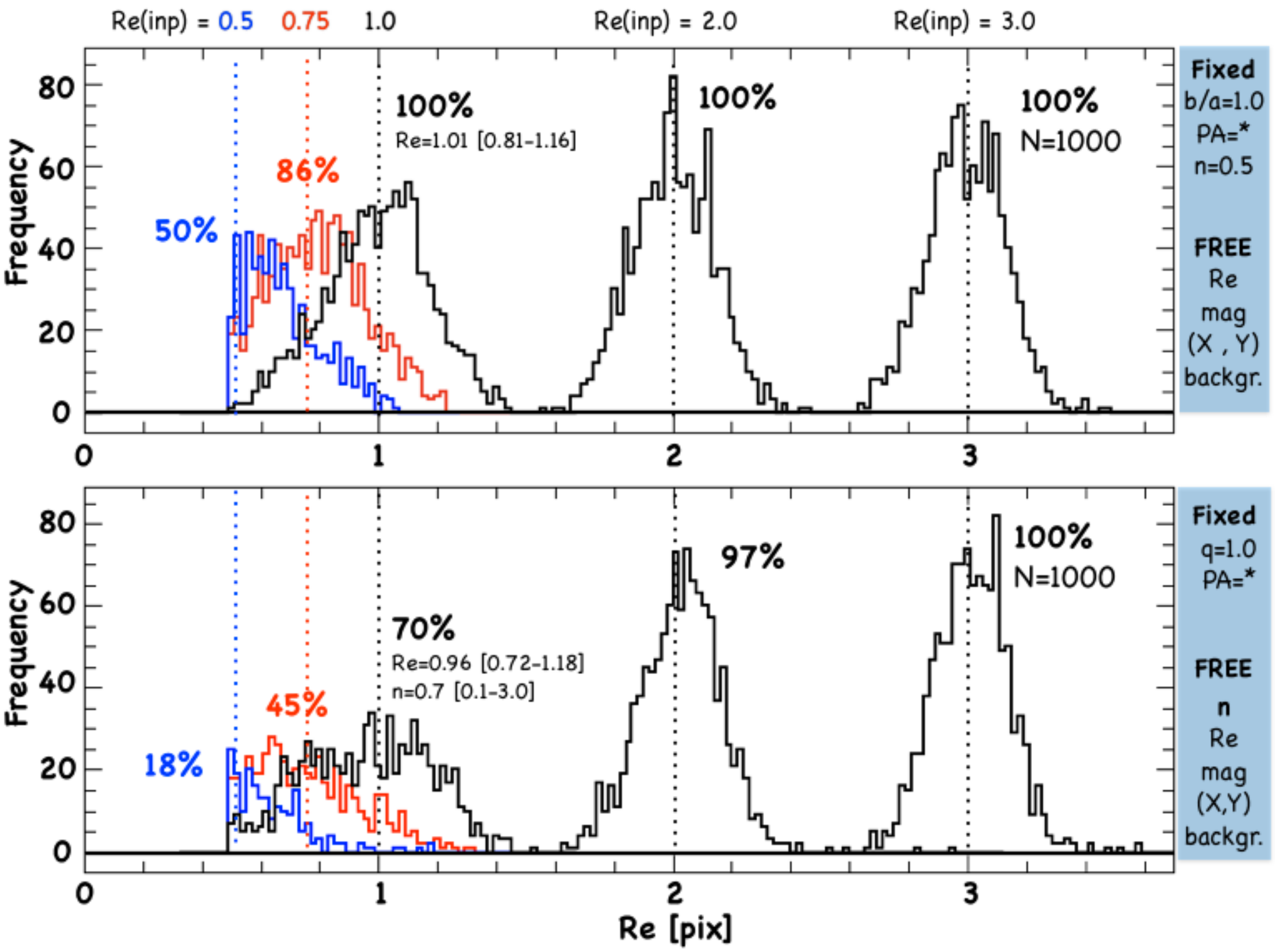} %Fig_MC_FREE.pdf} %Fig_GALFIT.pdf}
\caption{{\tt Galfit} fitting on the 1000 mock images allowing for its internal minimisation procedure. 
When a {\tt Galfit} warning was obtained (corresponding to  $R_e <0.5$ pix), the fit was not considered.  
The top panel shows the recovered $R_e$
for five sets of 1000 images, with $R_e = 3, 2, 1, 0.75$ and 0.5 pix, fixing  PA, q and $n=0.5$  (to the input values).
In the bottom panel PA and q are kept fixed (see the legend in the right side). 
Images with $R_e = 1$ pix are fully recovered, although with a wider spread. At sub-pixel scales the number of failures
increases (warnings on $R_e$ and/or $n$), however, the median still approaches the true value. This test is fully consistent with the results shown in
Figure~\ref{galfit_zoom}, suggesting the $R_e$ of D1(core) is less than 1 pix (13 pc).
}
\label{MC}
\end{figure}

\section{SED fitting of D1(core)}
\label{degene_core}

 \begin{figure*}
\centering
\includegraphics[width=17cm]{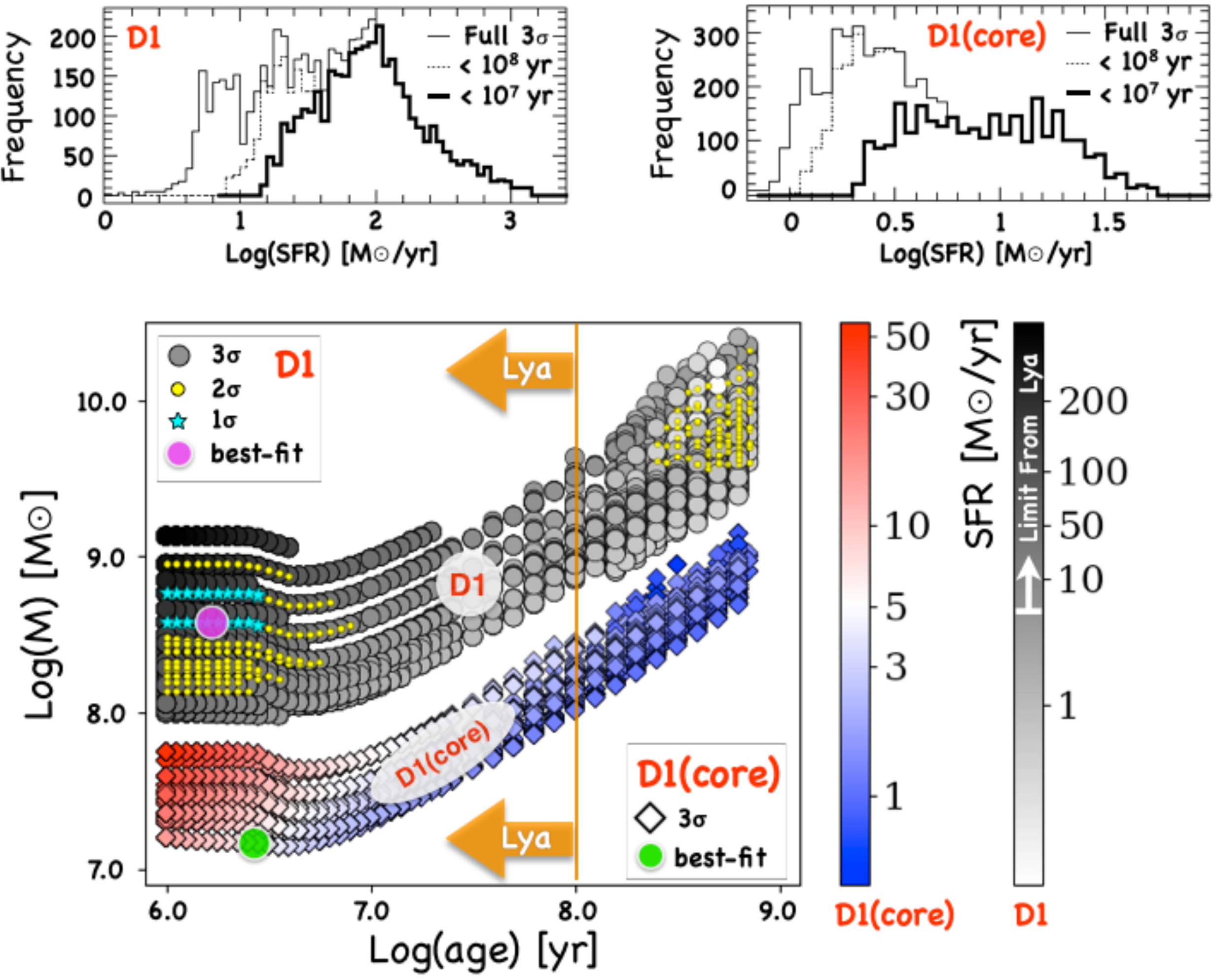} %FIG_degene_appendix.pdf}
\caption{The same as Figure~\ref{degene} in which the results within $3\sigma$ of the SED$-$fitting on D1 (gray coded) and D1(core) (colour-coded) are shown. In the top panels 
the 3$\sigma$ distributions of the SFRs are shown, including those with selected upper ages of 10 and 100 Myr. The magenta and green circles mark the best$-$fit soultions
for D1 and D1(core), respectively. }
\label{degeneall}
\end{figure*}

As discussed in Sect.~\ref{superdense} the SED fitting has been performed for the object D1 and the SFR of D1(core) was inferred by rescaling properly the
results from D1. Here we show the resulting physical properties from the SED$-$fitting applied directly to the extracted photometry of the core, excluding 
the Ks and the Spitzer/IRAC bands, not informative in this process.
The degeneracy among the stellar mass, age and star formation rate is evident (Figure~\ref{degeneall}), in which both D1 and D1(core) follow
a similar behaviour. The current spatial resolution in the MUSE data cube prevents us from measuring the \lya\ flux  separating among
D1 and D1(core). Therefore, as discussed in the main text (Sect.~\ref{lya}) a limit on the age and SFR can be obtaind only for D1. 
The best-fit solution for D1(core) suggests a stellar mass of $\simeq 0.8 \times 10^{6}$ \msun, an age younger than 10 Myr and a SFR  
of 0.35 \msunyr, the latter spanning the 3$\sigma$ range of $0.06-3.5$ \msunyr\  (the stellar mass
and the SFR shown in Figure~\ref{degeneall} are observed quantities, the intrinsic ones are obtainable by dividing the observed ones by $\mu_{tot} =17.4$).
Adopting the best estimate of the SFR and the upper limit on the effective radius of the core (Sect.~\ref{magD1}),  Log$_{10}(\Sigma_{SFR})$ turns out to be higher 
than 2.4, fully consistent with the distribution shown in Figure~\ref{SK},  lying in the upper part of the diagram populated by the densest known star-forming objects,
including young massive star clusters and ultra compact dwarf galaxies.

\end{document}